\documentclass[superscriptaddress, twocolumn, nobibnotes, prl]{revtex4-2}
\usepackage[T1]{fontenc}
\usepackage{libertine}
\usepackage{color, amsfonts, amsmath, graphicx, tikz, amssymb}
\usepackage{tabularx}
\usepackage{amsthm}
\usepackage{chemarr}
\usepackage[libertine, cmintegrals, bigdelims, vvarbb]{newtxmath}
\usepackage{appendix}
\usepackage{comment}
\definecolor{linkcolor}{rgb}{0,0,0.6} 
\usepackage{array,float}
\usepackage[pdftex,
colorlinks   = true,
pdfstartview = FitV,
linkcolor    = linkcolor,
citecolor    = linkcolor,
urlcolor     = linkcolor,
hyperindex   = true,
hyperfigures = false]
{hyperref}

\definecolor{lgreen} {RGB}{180,210,100}
\definecolor{dblue}  {RGB}{20,66,129}
\definecolor{jblue}  {RGB}{20,50,100}
\definecolor{nblue}  {RGB}{0,120,200}
\definecolor{dgreen} {RGB}{78,138,21}
\definecolor{ngreen} {RGB}{98,158,31}
\definecolor{lred}   {RGB}{220,0,0}
\definecolor{nred}   {RGB}{224,0,0}
\usepackage{cleveref}
\hypersetup{colorlinks,
            citecolor=blue,  
            filecolor=green,  
            linkcolor=red,  
            urlcolor=blue }

\usepackage{orcidlink}

\begin{document}

\title{Minimum action principle for entropy production rate of far--from--equilibrium systems}
\author{Atul Tanaji Mohite\,\orcidlink{0009-0004-0059-1127}}
\email{atul.mohite@uni-saarland.de}
\affiliation{Department of Theoretical Physics and Center for Biophysics, Saarland University, Saarbrücken, Germany}

\author{Heiko Rieger\,\orcidlink{0000-0003-0205-3678}}
\affiliation{Department of Theoretical Physics and Center for Biophysics, Saarland University, Saarbrücken, Germany}

\begin{abstract}
The Boltzmann distribution connects the energetics of an equilibrium system with its statistical properties, and it is desirable to have a similar principle for non-equilibrium systems. Here, we derive a variational principle for the entropy production rate (EPR) of far-from-equilibrium (fEQ) discrete state systems, relating it to the action for the transition probability measure of discrete state processes. This principle leads to a tighter, nonquadratic formulation of the dissipation function, speed limits, the thermodynamic-kinetic uncertainty relation, the large deviation rate functional, and the fluctuation relation, all within a unified framework of the thermodynamic length. Additionally, the optimal control of non-conservative transition affinities using the underlying geodesic structure is explored, and the corresponding slow--driving and finite--time optimal driving exact protocols are analytically computed. We demonstrate that discontinuous endpoint jumps in optimal protocols (defined as \textit{kinks}) are a generic, model--independent physical mechanism that reduces entropy production (EP) during finite--time driving of fEQ systems: `Geometric thermodynamic far-from-equilibrium shortcuts in swift state-to-state transformations'.
\end{abstract}

\date{\today}

\maketitle

{\textit{Introduction}}. \textemdash \: Stochastic Thermodynamics (ST) has emerged as a powerful framework for studying far-from-equilibrium (fEQ) systems, particularly finite-size systems prone to fluctuations~\cite{seifert_2012, sekimoto, Shiraishi_2023_book,ATM_2024_nr_st}. In this context, the entropy production rate (EPR) quantifies the thermodynamic dissipation required to sustain these out-of-equilibrium systems~\cite{seifert_2012,sekimoto,Shiraishi_2023_book,ATM_2024_nr_st}. ST has revealed fundamental laws of physics, including the Fluctuation Relation (FR), which captures the time-reversal asymmetry of the system~\cite{seifert_2012, sekimoto, Shiraishi_2023_book, ATM_2024_nr_st, schnakenberg_1976, Bochkov_1977, Bochkov_1979, Evans_1993, Evans_1994, Jarzynski_1997, Jarzynski_1997_pre, Crooks_1999, Tasaki_2000, Crooks_2000, Maes_2003, Gallavotti_1995, Lebowitz_1999, Kurchan_1998, Sekimoto_1997, Sekimoto_1998, Seifert_2005}, the Thermodynamic-Kinetic Uncertainty Relation (TKUR)~\cite{Gilmore_1985, Uffink_1999, Horowitz_2020, Barato_2015, Gingrich_2016, Horowitz_2017,Vo_2022,Kwon_2023_TUR_unified}, and Speed Limits (SL)~\cite{Ito_2018,Vo_2020,Lee_2022}, which together describe the trade-offs between precision, fluctuations, and dissipation in systems far from equilibrium. However, the relationship between TKUR and FR remains unclear, and they have been understood as distinct, independent fundamental laws in ST.

In equilibrium, a system maximizes its Gibbs entropy under given physical constraints (e.g., energy, Gibbs free energy)~\cite{Touchette_2009}. Non-equilibrium generalizations of this concept have been explored extensively~\cite{Klein_1954, Callen_1957, Glansdorff_1971, Jaynes_1980, Struchtrup_1998, Qian_2002, Evans_2004, Evans_2005, Lecomte_2005, Martyushev_2006, Wang_2006, Bruers_2007, Bruers_2007_maxEP_minEP, Lecomte_2007, Yoshida_2008, Martyushev_2010, Baule_2010, Niven_2010, Kawazura_2010, Doi_2011, Monthus_2011, Kawazura_2012, Chetrite_2013, Presse_2013, Endres_2017}, some with biological implications~\cite{Bialek_2000, Sasai_2003, Lan_2006, Vellela_2009}. However, a coherent structural understanding of fEQ systems remains elusive. Notably, most studies have focused on Gaussian state--space fluctuations, characterized by a quadratic dependence of the EPR on the driving force~\cite{Prados_2011, Bertini_2015, Bodineau_2004, Derrida_2007, Qian_2020}, valid under a close-to-equilibrium (cEQ) assumption~\cite{Onsager_1953, Onsager_1953_2}, but non-Gaussian fluctuations~\cite{Baule_2023,Huang_2025} become crucial in mesoscopic biological systems, where particle numbers are small and fluctuations are strong.
{In addition, existing state--space formulations fail to distinguish between the different transition channels to which fEQ systems are prone; therefore, the dynamics of fEQ systems cannot be fully captured by the state--space representation sufficient for equilibrium and cEQ systems. These shortcomings stem from the absence of an exact \textit{bottom-up approach} to non-Gaussian fluctuations and driving in transition--space.}

Recent advances have connected ST to the mathematical framework of Information Geometry (IG) \cite{Dechant_2018_multidimensional,Hasegawa_2019_pre,Van_Vu_2020_tur,VanVu_2021,kolchinsky_2022_information_geometry_epr,Ito_2020,Ito_2018,Ito_2024_omtp_st,Amari_2016_book}. 
However, the IG formulation compares different models in the control parameter space to compute the EPR, a statistical distance measure, namely the Kullback-Leibler (KL) divergence between the forward and reverse process. Due to this, it is sensitive to the correct or incorrect identification of the conjugate process. Moreover, it is also sensitive to the resolution of the trajectory by identifying all microscopic transitions. Thus, despite its precise mathematical formulation, its physical interpretation must be reconsidered when the identification of the backward process is not trivial, which is usually the experimental constraint. 

\begin{figure*}[t!]
\centering
\includegraphics[width=0.96\textwidth]{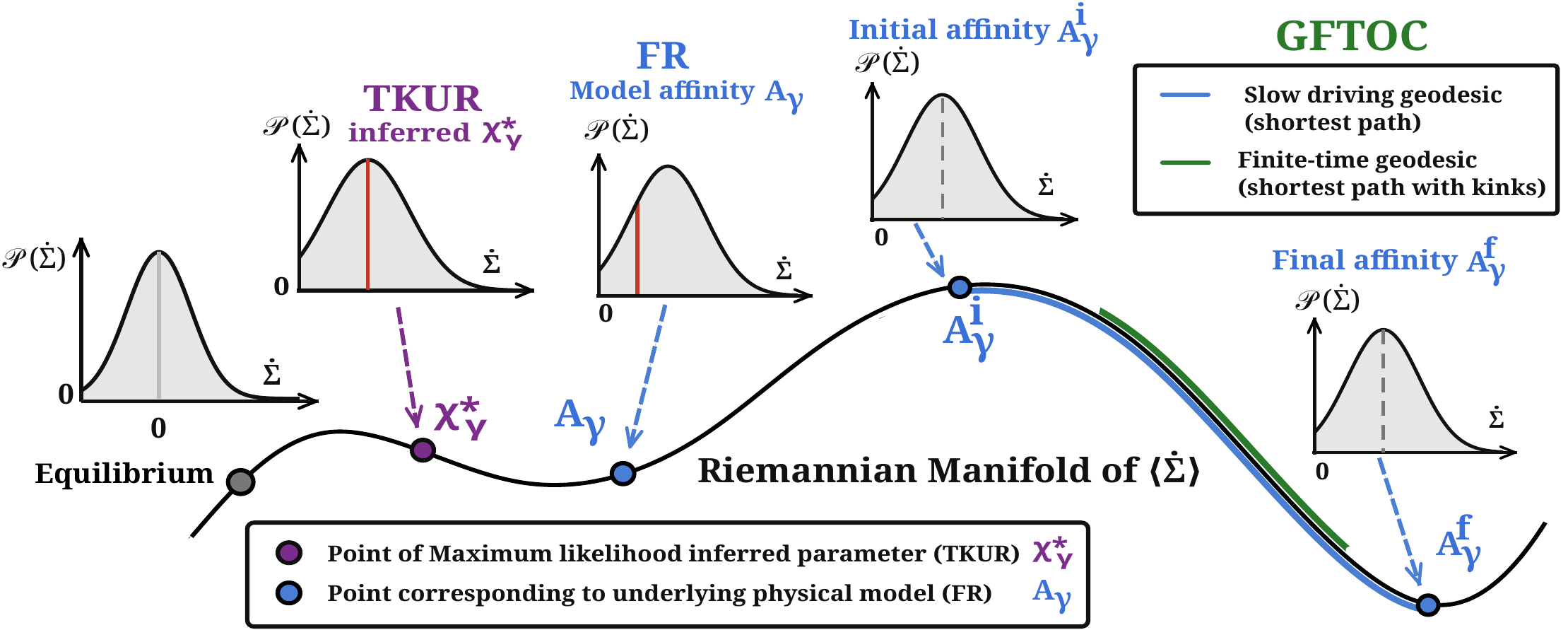}\label{fig:1}
\caption{ \textbf{Threefold geometric manifestation of MinAP:} The Riemannian manifold of the mean EPR $\langle \dot{\Sigma} \rangle$ is parameterized by $\{A_\gamma, D_\gamma\}$, and probability densities for $\dot{\Sigma}$. For a known model affinity $A_\gamma$, FR characterize the statistical properties of the stochastic $J_\gamma, T_\gamma$ (red vertical line denotes the observed value); EP is the most widely studied case. Agnostic of $A_\gamma$, \cref{eq:doi_peliti_path_integral_measure,eq:doi_peliti_transition_lagrangian} formulate a stochastic path-integral description of stochastic transition currents of the graph. Its variational solution yields the \textit{inferred} affinity $\chi_\gamma^* \neq A_{\gamma}$ (the most likelihood value: the red vertical line), along with the short--time and finite--time nonquadratic TKUR given respectively by \cref{eq:onsager_Machlup_functional,eq:EP_and_thermodynamic_length}. In the large-$\tau$ limit, the large deviation principle ensures that $\chi_\gamma^*$ converges to the \textit{true} physical affinity $A_\gamma$ of the underlying model. GFTOC addresses the problem of time-dependent driving a model affinity from $A_{\gamma}^i$ to $A_{\gamma}^f$ in a finite time $\tau$. Under the slow--driving constraint, the geodesic $\mathcal{G}(A_\gamma)$ gives the dissipation-minimizing optimal path (the cyan line connecting $A_\gamma^i$ and $A_\gamma^f$). However, the finite-$\tau$ makes infeasible to follow the geodesic exactly; this is circumvented by \textit{kinks} --- a `geometric shortcut', thereby a sub-arc of geodesic is traversed (green line). }
\label{fig:2}
\end{figure*}

In this work, we systematically derive a minimum action principle (MinAP) for the EPR of discrete state processes using Doi--Peliti field theory (DPFT) \cite{Doi_1976,Peliti,Weber_2017,ATM_2024_nr_st,ATM_2024_nr_cg,atm_2025_var_epr_derivation}. DPFT captures non-Gaussian transition fluctuations without the cEQ approximation: `the \textit{bottom--up} approach' \cite{ATM_2024_nr_st,ATM_2024_nr_cg,atm_2025_var_epr_derivation}. The resulting variational formulation provides a unified description of ST, incorporating nonquadratic TKUR, SL, and FR via Thermodynamic Length (TL). Building on this, we solve an optimal control problem for quasistatic and finite--time driving of non-conservative affinities controlling the housekeeping EPR of a unicyclic graph, a sub-case of generalized finite--time optimal control (GFTOC) \cite{atm_2025_gftoc}. We further prove an equivalence between this variational formulation and IG, extending IG methodologies in ST with a statistical-physical interpretation. 
We focus on the threefold manifestation of MinAP: nonquadratic TKUR, FR, and GFTOC (see \cref{fig:2}).

{\textit{Setup}}. \textemdash \: We consider a graph for Markov jump processes (MJPs) representing
the probability transport between microstates $i$ of a system
or equivalently linear chemical reaction networks (lCRNs),
$\rho_i$ denotes the probability density of state $i$ in MJPs
or the number and density of particles in lCRNs, and $\{i\}$ the set of all states.
$j_\gamma$ and $\chi_\gamma$ denote current and conjugate field for the generator of the transition $\gamma$ between two states. $\chi_\gamma$ characterizes the effective driving due to the stochastic fluctuation corresponding to the transition $\gamma$. $\{ \gamma^{\rightharpoonup} \}$ and $\{ \gamma^{\rightleftharpoons} \}$ denote the set of all unidirectional and bidirectional transitions of the graph. The transitions satisfy the Local Detailed Balance (LDB) condition via the transition affinity $ A_{\gamma} = \ln{(j_{\gamma}/j_{-\gamma})} = F_\gamma - \Delta_{\gamma} E + \Delta_{\gamma} S^{state} $ \footnote{$E$ and $F_\gamma$ are measured in units of the inverse temperature $\beta$,
where we set $k_b = 1$.}, which is composed of an external non-conservative driving $F_{\gamma}$ (which also serves as a control parameter, in addition to the control parameters of $E$), the change in the functional equilibrium energy $\Delta_{\gamma}E$ and the change $\Delta_\gamma S^{state}$ in the state entropy $S_{i}^{state} = -\ln{\rho_i}$ \cite{seifert_2012}. We define the total current and the traffic for $\gamma^{\rightleftharpoons}$, $J_\gamma = j_\gamma - j_{-\gamma}$, and $T_\gamma = j_\gamma + j_{-\gamma}$, respectively, which is a linearly independent decomposition of currents into antisymmetric and symmetric parts. The scaled $T_\gamma$ also characterizes the variance of the current \cite{ATM_2024_nr_cg,ATM_2024_nr_st,Maes_2020}, 
\footnote{The traffic scaled with a large deviation parameter characterizes the variance. For dynamical systems, the observation time $\tau$ \cite{Maes_2017,Maes_2020,Touchette_2009} 
is the relevant large deviation scaling parameter
and in fluctuating hydrodynamics \cite{Bertini_2015,Touchette_2009,ATM_2024_nr_cg,ATM_2024_nr_st} it is the system volume $\mathcal{V}$.}. The mobility for transition $\gamma^\rightleftharpoons$ is defined as $D_\gamma = \sqrt{j_\gamma j_{-\gamma}}$. Hence, $J_\gamma = 2D_\gamma \sinh{(A_\gamma/2)}$ and $T_\gamma = 2D_\gamma \cosh{(A_\gamma/2)}$ \cite{ATM_2024_nr_st,ATM_2024_nr_cg}. $\{D_\gamma, A_\gamma\}$ and $\{J_\gamma, T_\gamma\}$ formulate two equivalent descriptions of system dynamics, corresponding to `full' control and inference formulations, respectively. The stoichiometry matrix $\mathbb{S}$ manifests a contraction from $\{ \gamma^{\rightleftharpoons} \}$ to $\{i\}$ \cite{schnakenberg_1976}, through the continuity equation $\partial_t \vec{\rho} = \mathbb{S} \vec{J}$. 
%
%
{ Therefore, throughout this work, we use the transition--space representation for fEQ systems.}
The mean EPR for MJPs then has the bilinear form $\langle \dot{\Sigma} \rangle = \sum_{\{\gamma^\rightleftharpoons\}} \langle J_\gamma A_\gamma \rangle$ \cite{schnakenberg_1976}. 

{\textit{Variational formulation and most likelihood path}}. \textemdash \:
We derive an exact transition probability measure for the stochastic dynamics on graphs using DPFT \cite{atm_2025_var_epr_derivation}. It reads \footnote{ by construction, the transition probability measure is assumed to satisfy the normalization constraint, $1 = \int \mathcal{P} [\{J_\gamma, T_{\gamma}, \chi_\gamma \}] \mathbb{ D } \{ J_{\gamma} \} \mathbb{ D } \{ T_{\gamma} \} \mathbb{ D } \{ \chi_\gamma \} $. where $\mathbb{ D } \{ J_{\gamma} \} \mathbb{ D } \{ T_{\gamma} \}$ and $\mathbb{ D } \{ \chi_\gamma \}$ 
denote the path integral over all transition currents and noise realizations, respectively. }:
\begin{equation}
\mathcal{P} [\{J_\gamma, T_{\gamma}; \chi_\gamma \}]
= \exp\left(  -\int_{t_i}^{t_f} dt \;
\mathcal{L} \left[ \{ J_{\gamma}, T_\gamma; \chi_{\gamma} \} \right]
\right),
\label{eq:doi_peliti_path_integral_measure}
\end{equation}
where, $\mathcal{L}$ is the mesoscopic Lagrangian of the 
Doi-Peliti action $\mathcal{S}=\int dt \, \mathcal{L}$. The exact expression for $\mathcal{L}$ reads \cite{atm_2025_var_epr_derivation}:
\begin{equation}\label{eq:doi_peliti_transition_lagrangian}
    \mathcal{L}  \left[ \{ J_\gamma, T_\gamma; \chi_\gamma \} \right]
    = \sum_{\{\gamma^{\rightleftharpoons}\}} 
    \left[ 
    J_{\gamma} \left( \chi_\gamma + \sinh{ \left( \chi_\gamma \right) } \right) + T_{\gamma} 
    \left( 1 - \cosh{ \left( \chi_\gamma \right) }\right)
    \right].
\end{equation}
Remarkably, $\mathcal{L}$ incorporates all the higher cumulants of the current using $J_\gamma$ and $T_\gamma$ only. The saddle-point of $\mathcal{L}$ dominates the transition probability measure \cref{eq:doi_peliti_path_integral_measure} \cite{atm_2025_var_epr_derivation}. Solving the variational problem gives the optimal effective affinity $\chi_{\gamma}^* = 2 \tanh^{-1}{\left(J_{\gamma} / T_{\gamma} \right)}$, and the effective Lagrangian \cite{atm_2025_var_epr_derivation}: 
\begin{equation}\label{eq:onsager_Machlup_functional}
    \mathcal{L}^*[\{ J_{\gamma}, T_{\gamma} \}]   
    = \sum_{ \{ \gamma^{\rightleftharpoons} \} } 2 J_\gamma \tanh^{-1}{ \left( \frac{J_\gamma}{T_\gamma} \right) }.
\end{equation}
Physically, $\chi_{\gamma}^*$ corresponds to the most likely transition affinity that generates the given stochastic realization of $J_\gamma$ and $T_\gamma$;
see \cref{fig:1}\textcolor{red}{(a)}. Interpreted differently, it gives the effective transition affinity for the stochastic transition and quantifies the non-equilibrium-ness of $J_{\gamma}$ \cite{Andrieux_2007,Andrieux_2007_single_current_FT}. Moreover, $\chi_\gamma^*$ is inferred using current precision, $x = J_\gamma / T_\gamma$, highlighting its experimental relevance for thermodynamic inference. The current precision is a quantitative physical measure of the system's non-equilibrium-ness, the further the system is from equilibrium, the larger the value of $x$. Importantly, the non-linear dependence of $\chi_{\gamma}^*$ on the current precision is attributed to effectively incorporating all higher-order current cumulants, ensuring a robust framework for small-size fEQ systems prone to non-Gaussian fluctuations \cite{atm_2025_var_epr_derivation}. The Gaussian or cEQ approximation implies $\tanh^{-1}{(x)} \approx x$, however, $\tanh^{-1}{(x)} \geq x$, for which reason the mismatch is more profound for fEQ systems. This nonlinearity will play a key role throughout this paper.

If the transition affinities were known, then $\mathcal{L}^*$ 
would trivially equal the mean EPR, since $ \tanh{(A_\gamma/2)} = J_\gamma/T_\gamma $ and therefore $ \mathcal{L}^* = \sum_{\{\gamma^\rightleftharpoons\}} J_\gamma A_\gamma$, valid for the most likelihood path. However, $\mathcal{L}^*$ defines the EPR using the current and traffic and therefore corresponds to the inferred mean EPR, valid even for a less likelihood path, for which $\chi_\gamma^* \neq A_\gamma$. 
{For example, in \cref{fig:1}\textcolor{red}{(a)}, an observed $\chi_\gamma^*$ (vertical dotted dark orange line) could emerge from the most likelyhood path of a model (orange line, for which $\chi_\gamma^* = A_\gamma$) or a less likelyhood path of another model (cyan line, for which $\chi_\gamma^* \neq A_\gamma$).}
Notably, $\mathcal{L}$ and $\mathcal{L}^*$ have been referred to as information-geometric EPR in \cite{kolchinsky_2022_information_geometry_epr}. In contrast to \cite{kolchinsky_2022_information_geometry_epr}, here the EPR is defined by using the control parameter of the model itself and does not require the identification of the conjugate process or statistics. Hence, our formulation addresses the issue associated with the physical interpretation of IG and proves its equivalence with the statistical mechanical formulation of ST. $\mathcal{L}^*$ is an exact nonquadratic dissipation function, in contrast to the cEQ quadratic Onsager-Machlup functional \cite{Onsager_1953,Onsager_1953_2}. The set of \cref{eq:doi_peliti_path_integral_measure,eq:onsager_Machlup_functional}, 
our first main result, 
formulates a min-max variational formulation for the action, in particular, 
$ \Sigma = \inf_{\{J_{\gamma}, T_{\gamma} \}} \sup_{ \{\chi_{\gamma} \} } \mathcal{S} \left[ \left\{ J_{\gamma}, T_{\gamma} ; \chi_{\gamma} \right\} \right] $, which physically corresponds to MinAP valid for fEQ systems \cite{atm_2025_var_epr_derivation}. 
\begin{figure*}[t!]
\centering
\includegraphics[width=\textwidth]{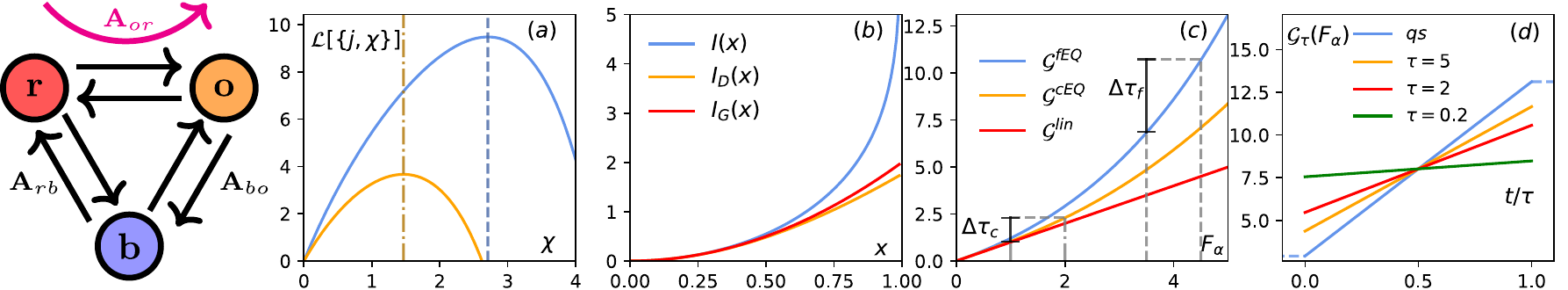}\label{fig:1}
\caption{ (Leftmost panel) An illustration for a three-state (red-orange-blue) unicyclic graph. The magenta curved arrow denotes the control of the non-conservative part $F$ of the cyclic transition affinity $A_{or} + A_{bo} + A_{rb}$ through a coupling to an external enzymatic reservoir. (a) Lagrangian $\mathcal{L}[\{J,T;\chi\}]$ [\cref{eq:doi_peliti_transition_lagrangian}] for fixed $J_\gamma=3.5, T_\gamma=4$ (cyan) and $J_\gamma=2.5, T_\gamma=4$ (orange). The corresponding most likelihood transition affinity $\chi^*_\gamma$ as a vertical dotted lines. (b) Comparison between exact $I = 2x\tanh^{-1}(x)$, dynamical $I_{D}= 2x\sinh^{-1}(x)$ and Gaussian $I_{G} = 2x^2$  rate functional, where, $x=J_\gamma/T_\gamma$ is the current precision. (c) $\mathcal{G}(F): F \to t$.  Comparison between cEQ geodesic [ $\mathcal{G}^{cEQ}(F)$ ], fEQ geodesic [$\mathcal{G}^{fEQ}(F)$ ] and linear geodesic [$\mathcal{G}^{lin}(F)$]. For the fixed $v_{qs}$, the $F^f - F^i = 1$ is considered for the cEQ $F^i=1$ and fEQ $F^i = 3.5$ and the corresponding $\Delta \tau_c$ and $\Delta \tau_f$ are plotted, which are the minimum driving time required. (d) The finite--time optimal protocol $\mathcal{G}_{\tau}$ is plotted for the different values of $\tau$ with the same initial and final value condition (shown by the dotted blue lines). }
\label{fig:1}
\end{figure*}

{\textit{Thermodynamic length and Entropy production}}. \textemdash \:
We define the thermodynamic length of a transition $\gamma^\rightleftharpoons$ as $\tau \tilde{J}_{\gamma} = \int_{0}^{\tau} J_{\gamma}$ and $ \tau \tilde{T}_{\gamma} = \int_{0}^{\tau} T_{\gamma} $, referring to time-integrated current and traffic. 
{$\tilde{x}_\gamma = \tilde{J_\gamma}/\tilde{T}_\gamma$ defines the time-integrated current precision.}
By integrating \cref{eq:onsager_Machlup_functional} from time $t=0$ to $t=\tau$, the relationship between thermodynamic length and EP $\Sigma$ is,
\begin{equation}\label{eq:EP_and_thermodynamic_length}
    \tau \tilde{\Sigma} = \Sigma = \int_0^{\tau} \mathcal{L}^* d t \geq \sum_{ \{\gamma^{\rightleftharpoons}\} } 2 \tau \tilde{J}_{\gamma} \tanh^{-1}{ \left( \frac{ \tilde{J}_{\gamma} }{ \tilde{T}_{\gamma} } \right) },
\end{equation}
representing a nonquadratic fEQ formulation of thermodynamic length, valid beyond the existing cEQ formulations \cite{Salamon_1983,Salamon_1985,Schlogl_1985,Brody_1995,Crooks_2007}. It delineates the trade-off between current length, fluctuations, and EP. Due to \cref{eq:doi_peliti_path_integral_measure}, it connects $\tilde{J}_{\gamma}$ and $\tilde{T}_{\gamma}$ to the transition probability measure 
{through the EP contribution $\Sigma_\gamma$ due to $\gamma^\rightleftharpoons$.} 
Hence, \cref{eq:onsager_Machlup_functional,eq:EP_and_thermodynamic_length} correspond to the short--time TL ($\tau \to 0$) and finite--time TL, respectively. Defining $I(x) = 2x \tanh^{-1}{(x)}$ and it's inverse $I^{-1}(x)$. We invert \cref{eq:EP_and_thermodynamic_length} to obtain the nonquadratic bounds using TL:
\begin{equation}\label{eq:precision_EPR_relation}
    \tilde{J}_{\gamma} \leq \tilde{T}_{\gamma} I^{-1}\left( \frac{\Sigma_{\gamma}}{\tau \tilde{T}_{\gamma}} \right) \leq \sqrt{ \frac{\Sigma_{\gamma} \tilde{T}_{\gamma}}{2 \tau} },
\end{equation}
since $I(x) \geq 2x^2$, giving the tightest exact upper bound on current precision than quadratic counterparts corresponding to TKUR \cite{Lee_2022}. Our second main result \cref{eq:EP_and_thermodynamic_length,eq:precision_EPR_relation} formulates a connection between thermodynamic length and $\Sigma_\gamma$ for fEQ systems. 

\textit{The exact large deviation rate functional}. \textemdash \:
The set of \cref{eq:doi_peliti_path_integral_measure,eq:onsager_Machlup_functional,eq:EP_and_thermodynamic_length} implies, $\mathcal{P}\left[ \{ \tilde{J}_{\gamma}, \tilde{T}_\gamma \} \right] \asymp e^{-\tau \sum_{ \{ \gamma^\rightleftharpoons \} } \tilde{T_\gamma} I(\tilde{x}_\gamma) }$. Notably, by replacing $\tilde{\Sigma} \to E$ and $\tau \to \beta$, the canonical ensemble analogue for fEQ systems is identified using dynamical physical quantities, scaled time-integrated currents and traffics here. Such non-equilibrium canonical ensemble analogues are known in the Large deviation theory, and $I$ is known as the rate functional \cite{Touchette_2009}. However, previous studies have obtained quadratic $I_G(x) = 2x^2$ and nonquadratic $I_D(x) = 2x\sinh^{-1}{(x)}$ rate functions using the Gaussian approximation and the non-equilibrium fluctuation-response relation, respectively \cite{atm_2025_var_epr_derivation,Maes_2008,Maes_2009,Maes_2017,Mielke_2014_ldp,kobayashi_2023_information_graphs_hypergraphs}. We observe a profound mismatch between $I, I_G$ and $I_D$ for fEQ systems, see \cref{fig:1}\textcolor{red}{(b)}. Physically, this implies that the established rate functionals massively underestimate $\Sigma$ for fEQ systems. In contrast, our third main result, the `exact' rate functional $I$ avoids this problem due to `the \textit{bottom--up} approach' and gives the tightest and exact bounds on $\Sigma$ \cite{atm_2025_var_epr_derivation}.

{ {\textit{MinAP for fEQ systems beyond ST}}. \textemdash \:
LDB in ST relies on two assumptions: timescale separation and weak coupling between the system and environment \cite{seifert_2012,sekimoto,Shiraishi_2023_book}. Since $\chi_\gamma$ and $A_\gamma$ define the stochastic transition parameters and the thermodynamically consistent model (LDB), respectively, relaxing both of the above-mentioned assumptions, MinAP [\cref{eq:doi_peliti_path_integral_measure,eq:doi_peliti_transition_lagrangian,eq:onsager_Machlup_functional}] applies to any fEQ system beyond ST, regardless of whether LDB holds. However, identifying $\mathcal{L}^*$ with $\dot{\Sigma}$ requires thermodynamic consistency imposed by LDB; without it, $\mathcal{L}^*$ should be interpreted in an information-geometric sense, retaining a physically grounded connection to transition currents and traffics \cite{atm_2025_var_epr_derivation}. The \textit{bottom--up} validity of LDB and ST emerges naturally as a sub-case of MinAP when $\chi_\gamma^*$ is time-homogeneous due to the decay of system-environment correlations over a sufficiently large timescale \cite{Evans_2004,Evans_2005,atm_2025_var_epr_derivation}.}

\textit{{MinAP under coarse-graining}}. \textemdash \:
Microscopic currents are not feasible to observe experimentally, but coarse-grained observable (macroscopic) currents are. Thus, naturally, we extend the formulations to observable currents. We define a set of observable (macroscopic) currents $\{J_o\}$ and traffic $\{T_o\}$. They are carefully chosen, so they respect the scaling of EPR and avoid double counting of microscopic transitions \cite{atm_2025_var_epr_derivation}. We derive Lagrangian $(\mathcal{L}_{\{ o\}}^*[\{ J_{o}, T_{o} \}])$, the inferred transition affinity $\left( \chi_o^* = 2 \tanh^{-1}{ \left( {J_o}/{T_o} \right) } \right)$ and  inferred EPR $(\dot{\Sigma}_{\{o\}} = \sum_{\{o\}} \chi_o^* J_o)$ for observable currents \cite{atm_2025_var_epr_derivation}, the relationship similar to \cref{eq:onsager_Machlup_functional} reads,
\begin{equation}\label{eq:onsager_Machlup_functional_cg_observable}
    \dot{\Sigma}_{\{o\}} = \mathcal{L}_{ \{o\} }^*[\{ J_{o}, T_{o} \}]   
    = \sum_{ \{o\} } 2 J_o \tanh^{-1}{ \left( \frac{J_o}{T_o} \right) }.
\end{equation}
Hence, analogously to \cref{eq:EP_and_thermodynamic_length}, the finite--time TL holds for observable currents. $ \mathcal{L}_{\{ o\}}^*$ gives a lower bound on $\mathcal{L}^*$ \cite{atm_2025_var_epr_derivation}. Physically, it corresponds to observable currents and traffics are able to capture a part of the microscopic EPR, $\dot{\Sigma} \geq \dot{\Sigma}_{\{o\}}$. Bounds on $\dot{\Sigma}$ are tightened by exploiting two factors. First, the `exact' nonquadratic rate functional, as discussed previously. Second, by selecting the correct (all microscopic linearly independent) observable currents, $\{o\} = \{\gamma^\rightleftharpoons\}$ \cite{atm_2025_var_epr_derivation}. 
{Importantly, \cref{eq:onsager_Machlup_functional_cg_observable} defines a `physical' inferred EPR from the time-antisymmetric and time-symmetric parts of physical observables, even when the underlying model violates LDB or being agnostic to microscopic thermodynamics.}

\textit{Nonquadratic state--space TKUR, SL and OM functional}. \textemdash \:
We consider three notable cases of observable currents, exhibiting the application of \cref{eq:onsager_Machlup_functional_cg_observable}. Case $(1)$: The observable vorticity currents $\omega_{ij} = \rho_i \partial_t \rho_j -\rho_j \partial_t \rho_i$ between states $\rho_i$ and $\rho_j$. Defining $C_{ij}(\tau) = \rho_i(\tau) \rho_j(0), C_{ij}^s(\tau) = \rho_i(\tau) \rho_j(0) + \rho_j(\tau) \rho_i(0)$ and $C_{ij}^a(\tau) = \rho_j(\tau) \rho_i(0) - \rho_i(\tau) \rho_j(0)$ quantifies the temporal state correlations, its symmetric and anti-symmetric components, respectively, and $\Delta_0^\tau C_{ij}^a(\tau) = C_{ij}^a(\tau) - C_{ij}^a(0)$ and $\Delta_0^\tau C_{ij}^s(\tau) = C_{ij}^s(\tau) - C_{ij}^s(0)$. The finite--time TL for $\omega_{ij}$ leads to the state--space nonquadratic TKUR: 
\begin{equation}\label{eq:inferred_EP_and_thermodynamic_length_correlation}
    {\Sigma}_{\{\omega\}} \geq \sum_{ \{ij\} } 2 \Delta_0^\tau {C}_{ij}^a \tanh^{-1}{ \left( \frac{ \Delta_0^\tau {C}_{ij}^a }{ \Delta_0^\tau {C}_{ij}^s } \right) },
\end{equation}
as was first obtained by us for microscopically nonreciprocal systems in Ref. \cite{ATM_2024_nr_st}, but is valid for an effective (emergent) nonreciprocal description of fEQ process, which is not necessarily microscopically nonreciprocal. The tightest possible lower bound on $\Sigma$ is obtained using $C_{ij}(\tau)$ and \cref{eq:inferred_EP_and_thermodynamic_length_correlation}. Importantly, the correlations $C_{ij}(\tau)$ are easily accessible experimentally \cite{Terlizzi_2024}, in contrast to the usual current-space formulation of the TKUR. This is our fourth main result. Case $(2)$: Total currents $(\{J_i\})$ into $\{ \rho_i \}$. The finite--time TL leads to the nonquadratic speed limit bound on the EP, $\Sigma_{SL} \geq 2 [ \rho_i(\tau) - \rho_i(0) ] \tanh^{-1}{ ( { \left[\rho_i(\tau) - \rho_i(0) \right] }/{ \tau \tilde{T}_{i} } ) }$. Quantifying the minimum EP needed to change the state distribution, $\{\rho_i(0) \} \to \{\rho_i(\tau)\}$. Case $(3)$: The observable relaxation currents $(\{J_i^{rel}\})$ to $\{\rho_i\}$, $ \mathcal{L}_{OM}^* = \sum_{\{i\}} 2 (\partial_t \rho_i - J_i^{ss}) \tanh^{-1}{ [ {(\partial_t \rho_i - J_i^{ss})}/{T_i^{rel}} ] }$. Equivalently, due to the relaxation-fluctuation symmetry, $\mathcal{L}_{OM}^*$ quantifies non-Gaussian fluctuations around steady-state, generalizing the Gaussian Onsager-Machlup functional derived around equilibrium ($J_i^{ss} = J_i^{eq}$) \cite{Onsager_1953,Onsager_1953_2}. 

{\textit{Fluctuation relation}}. \textemdash \: Using the bilinear form of the EPR $\dot{\Sigma}_{\{o\}} = \sum_{\{o\}} \chi_o^* J_o$, the normalization condition for $\mathcal{P}$ \cref{eq:doi_peliti_path_integral_measure,eq:EP_and_thermodynamic_length,eq:onsager_Machlup_functional_cg_observable}, the integral fluctuation relation for the inferred EP $\langle e^{-\tau \tilde{\Sigma}_{\{o\} } } \rangle = 1$, and for observable currents $\langle e^{-\tau \chi_o^{*} \tilde{J}_{o } }\rangle = 1$, are all satisfied. Our formulation reveals a fundamental, previously unknown connection between the FR and the nonquadratic TKUR, unified by the MinAP. The nonquadratic TKUR concerns the inference problem, whereas the FR formulates the corresponding control description, where the effective affinities are known and implications for non-equilibrium fluctuations of the observable current are studied. Hence, the FR and nonquadratic TKUR are two sides of the same coin -- our fifth main result. This also reveals the shortcomings of the quadratic TKUR (an approximate law). We find that stochastic EP is the most precise special observable current that maximizes the thermodynamic inference of the system's non-equilibrium-ness~\cite{atm_2025_var_epr_derivation}. Effective affinity is the key thermodynamic inference property that reveals the underlying Martingale property of microscopic and observable currents~\cite{Chetrite_2011,Neri_2017,atm_2025_var_epr_derivation}.

\textit{{Geodesic structure and optimal control of fEQ systems}}. \textemdash \: 
{An application of MinAP presented in \cite{atm_2025_gftoc} develops the GFTOC framework, resolving major problems identified in Ref.\cite{Guery_Odelin_2023},
{precisely defined as, the `thermodynamic optimization and interpretation' of the `geometric shortcuts' in the `swift state-to-state transformations of fEQ systems'.}
To illustrate this framework, we consider a unicyclic three-state graph (\cref{fig:1}, leftmost panel) representing an equivalence class of prototypical biological and chemical systems.
{Thus, $\{i\} = \{r,o,b\}, \{\gamma^{\rightleftharpoons}\} = \{rb,bo,or\}$ with control parameters $F_{\gamma} = F, -\Delta_\gamma E = 0,  \forall \gamma \in \{\gamma^{\rightleftharpoons}\}$.} 
We formulate a full control problem, 
{(`thermodynamic optimization') } for slow driving of $F$, which controls the housekeeping EPR $\dot{\Sigma}^{hk}  = 4F\sinh{(F/2)} = \mathcal{L}^*$; from $F^i$ to $F^f$ in the driving time $\tau$ 
{(a `state-to-state transformation')}. The system is assumed to relax instantaneously to the steady-state distribution, leading to vanishing excess EP and a control problem for the optimization of $\dot{\Sigma}^{hk}$. The corresponding `slow' driving Lagrangian then reads \cite{Salamon_1985,Brody_1995,Schlogl_1985,Crooks_2007,atm_2025_gftoc}:}
\begin{equation}\label{eq:optimal_driving_epr}
     \mathcal{L}_{drv}^*  \left[ F, \dot{F} \right] = \frac 1 2 \partial_{{F}}^2 \mathcal{L}^* \dot{F}^2,
\end{equation}
where, $\partial_{{F}}^2 \mathcal{L}^*$ is the local curvature of $\mathcal{L}^*$ with respect to $F$ and plays the role of mass in the driving 
\textit{kinetic energy} [\cref{eq:optimal_driving_epr}], 
{
in addition to \textit{potential energy} [\cref{eq:onsager_Machlup_functional}].
}
The fEQ systems exhibit higher mass due to stronger fluctuations and dissipation, $\partial_{{F}}^2 \mathcal{L}^* = T + \frac 1 4 \mathcal{L}^*$, and critical slowing from the singularity in the limit $F \to \infty$ \cite{atm_2025_gftoc}. The geodesic, $\mathcal{G}(F)$, denotes the optimal path that minimizes the total driving EP $(\Sigma_{qs} = \int_0^\tau \mathcal{L}_{drv}^* dt = \frac 1 2 \int_0^\tau [\partial_t \mathcal{G}(F)]^2 )$, and is defined as the linear interpolation between the initial and final state, $\mathcal{G}(F) = ({t}/{\tau})\,\mathcal{G}(F^f) + ( 1 - {t}/{\tau} ) \mathcal{G}(F^i)$.
{Using $\mathcal{G}(F)$ in $\Sigma_{qs}$, a `time-dissipation-geometry' tradeoff relation is unveiled,
\begin{equation}\label{eq:time_dissipation_geometry_tradeoff}
    \Sigma_{qs} = \frac{1}{2 \tau} [ \mathcal{G}(F^f) - \mathcal{G}(F^i) ]^2.
\end{equation}
It implies that the larger geometric distance traversed in control parameter space ($F$) has to be achieved through a tradeoff between larger driving EP $\Sigma_{qs}$ and driving time $\tau$.
}
The analytical form of $\mathcal{G}$ plays a crucial role [\cref{fig:1}\textcolor{red}{(c)}], for the three different regimes: linear, cEQ and fEQ optimal control. The tightness of the exact bound on $\Sigma_{qs}$ using $\mathcal{G}^{fEQ}$ is prominent for fEQ systems.
{For fixed $F^i, F^j$ and $\tau$, $\mathcal{G}^{fEQ} > \mathcal{G}^{cEQ} > \mathcal{G}^{lin}$ [\cref{fig:1}\textcolor{red}{(c)}] implies that prominent violations of \cref{eq:time_dissipation_geometry_tradeoff} using slow--driving constraint in fEQ systems, requiring careful treatment of the finite-$\tau$ constraint.}
GFTOC under slow-driving, derived in Ref.\cite{atm_2025_gftoc}, surpasses existing cEQ formulations that utilize optimal control theory \cite{Jordan_1998,Benamou_2000,Aurell_2011,Van_vu_2023} and thermodynamic geometry \cite{Salamon_1983,Salamon_1985,Schlogl_1985,Brody_1995,Crooks_2007,Sekimoto_1997_slow_driving_optimal_control,Sivak_2012,Mandal_2016,Li_2022},
 making it our sixth main result.

\textit{{Generalized finite--time optimal control}}. \textemdash \: The finite--time optimal protocol exhibits `kinks' \cite{Schmiedl_2007,Aurell_2011,Chen_2019_stochastic_control,Zhong_2024}, in contrast to the slow--driving approach \cite{Sekimoto_1997_slow_driving_optimal_control,Sivak_2012,Mandal_2016,Li_2022}. Within existing cEQ `model--specific' formulations \cite{Schmiedl_2007,Zhong_2024}, this mechanism has been physically interpreted as an artifact of boundary condition constraints. We examine the possibility of `kinks', and obtain the driving EP minimizing finite--time optimal protocol $[\mathcal{G}_{\tau}(F)]$ \cite{atm_2025_gftoc}. It is equivalent to substituting $t/\tau \to t/(\tau + 2)$, $1 \to {(1 + \tau)}/{(2+\tau)}$ and $0 \to {1}/{(2+\tau)}$ in $\mathcal{G}(F)$, plotted in \cref{fig:1}\textcolor{red}{(d)} for different $\tau$:
{a `swift state-to-state transformation'}. Physically, the optimal finite--time driving corresponds to total driving time $\tau + 2$ with initial and final times being $t=1$ and $t=\tau+1$, respectively, leading to `kinks': 
{a `geometric shortcut' required for `swift state-to-state transformation of fEQ systems', which restores the `time-dissipation-geometry' relation [\cref{eq:time_dissipation_geometry_tradeoff}] for any finite-$\tau$ slow--driving.} The analytical expression for $\mathcal{G}_{\tau}(F)$ reads:
\begin{equation}\label{eq:finite_time_optimal_protocol}
     \mathcal{G}_{\tau}(F) = \left( \frac{1+\tau}{2+\tau} - \frac{t}{2 + \tau} \right) \mathcal{G}(F^{i}) + \left( \frac{1}{2+\tau} + \frac{t}{2 + \tau} \right) \mathcal{G}(F^{f}).
\end{equation}
Importantly, this formulates a `model--independent geometric unification' of finite--time and slow--driving optimal processes and reveals fundamental aspects of thermodynamic control \cite{atm_2025_gftoc}, which is our seventh main result. For the fast-driving limit $\tau \to 0$, \cref{eq:finite_time_optimal_protocol} gives the midpoint interpolation in the geodesic space, $\mathcal{G}_{\tau}(F) =  [ \mathcal{G}(F^{i}) + \mathcal{G}(F^{f}) ]/2$, see \cref{fig:1}\textcolor{red}{(d)}.

Using \cref{eq:finite_time_optimal_protocol}, the finite--time driving EP reduces to $\Sigma_{\tau} = [ \mathcal{G}(F^f) -  \mathcal{G}(F^i) ]^2 /[2(2+\tau)]$ and is smaller than $\Sigma_{qs}$. This is realized through the physical mechanism named as a `thermodynamic shock' to the system \cite{atm_2025_gftoc}. Here, `thermodynamic shock' refers to the instantaneous global thermodynamic cost imposed on the system to generate `kinks'. 
{ Therefore, this `thermodynamically optimum' solution admits a `physical interpretation' analogous to the work-heat dual relation for any fEQ finite--time time-dependent driven process, where heat and work refer to the thermodynamic costs of `discontinuous kinks' and `continuous slow--driving' [\cref{eq:optimal_driving_epr}], respectively. }
The small $\tau$ divergence (underestimation) from $\Sigma_\tau \propto 1/\tau$ scaling has been experimentally observed \cite{Ma_2020}, implying that the mechanics of the `thermodynamic shock' is real and a quantitative check of our predictions would be worthwhile.
Our formulation of GFTOC delineates a general model--independent underlying principle for finite--time driving, presented in Ref.\cite{atm_2025_gftoc}. \Cref{table:threefold_manifestation_of_MinAP} and \cref{fig:2} summarizes the threefold manifestation of MinAP. 
\begin{table}[t!]
\centering
\begin{tabular}{c c c c}
     \hline \hline
     \textbf{Description} & \textbf{Thermodynamic} & \hspace{0.2cm} \textbf{Partial} & \hspace{0.2cm} \textbf{Full} \\[-0.1ex]
     \vspace{1pt} \textbf{level} & \textbf{inference} & \hspace{0.2cm} \textbf{control} & \hspace{0.2cm} \textbf{control} 
     \\ [0.2ex] 
     \hline
     Thermodynamic & Nonquadratic & & \\ [-0.0ex]
     \vspace{1pt} law & TKUR \cite{atm_2025_var_epr_derivation} & \hspace{0.2cm} FR \cite{atm_2025_var_epr_derivation}  & \hspace{0.2cm} GFTOC \cite{atm_2025_gftoc} \\ [-0.2ex]
     \hline
     {Controllable} &  & fixed  & time-dependent \\ [-0.0ex]
     \vspace{1pt} {parameters} & - & \hspace{0.2cm} $A_\gamma$  &  \hspace{0.2cm} $A_\gamma, D_\gamma$  \\ [0.1ex]
     \hline
     Observable &  &  &  \\ [-0.0ex]
     \vspace{1pt} quantities & $J_\gamma, T_\gamma$ & \hspace{0.2cm} $J_\gamma$ & \hspace{0.2cm} - \\ [0.2ex]
     \hline
     EPR Contribution & \textit{potential} & \textit{potential} & \textit{kinetic} \\ [-0.2ex]
     \vspace{1pt} to the action is & \textit{energetic} & \textit{energetic} & \textit{energetic} \\ [-0.3ex]
     \hline
     TL Contribution & Stochastic $J_\gamma, T_\gamma$  & Stochastic $J_\gamma$  & An explicit \\ [-0.3ex]
     \vspace{1pt} to the action & for fixed but  & at fixed and  & time-dependent \\ [-0.5ex]
     due to & unknown $A_{\gamma}$ & known $A_{\gamma}$ & change of $A_{\gamma}, D_\gamma$ \\ [-0.1ex]
     \hline \hline
\end{tabular}
\caption{Threefold geometric manifestation of MinAP}
\label{table:threefold_manifestation_of_MinAP}
\end{table}

\textit{Conclusion and Outlook}. \textemdash \:We present a unified MinAP framework for the EPR of discrete-state processes. Deriving the exact path integral representation incorporates non-Gaussian fluctuations, yielding a nonquadratic dissipation function. This gives a physical interpretation of the action Lagrangian as the inferred EPR, analogous to the energy functional in the equilibrium Boltzmann distribution. We further derive an exact nonquadratic large deviation rate functional, tightening EPR bounds relative to existing formulations. Our framework unifies the nonquadratic TKUR, SL, and FR via thermodynamic length, applicable to both microscopic and observable currents \cite{atm_2025_var_epr_derivation}. Building on this, we address the optimal control problem for fEQ systems, where the non-conservative affinity is driven between initial and final values. Identifying the corresponding geodesic solves this problem, incorporating finite--time effects to analytically compute optimal protocols and the associated housekeeping EPR cost. Our results reveal \textit{kinks} as a generic mechanism minimizing finite--time driving dissipation; here, we summarize this prototypical mechanism for fEQ systems. GFTOC nonetheless enables computationally feasible applications \cite{atm_2025_gftoc}, spanning experimental settings for the thermodynamically efficient design and control of fEQ biological systems and nano-materials in finite time. This opens the door to further practical applications of the MinAP. Extensions to nonlinear CRNs using hypergraphs are also straightforward \cite{kobayashi_2023_information_graphs_hypergraphs}, given the generality of our formulation. 

\section{acknowledgments}
ATM thanks Jin-Fu Chen for highlighting the experimental work \cite{Ma_2020}, which inspired the theoretical formulation of GFTOC \cite{atm_2025_gftoc}.

%
%
\vskip 0.5cm
\subsection*{References}
\bibliographystyle{apsrev4-1}
\bibliography{reference}

\begin{thebibliography}{117}%
\makeatletter
\providecommand \@ifxundefined [1]{%
 \@ifx{#1\undefined}
}%
\providecommand \@ifnum [1]{%
 \ifnum #1\expandafter \@firstoftwo
 \else \expandafter \@secondoftwo
 \fi
}%
\providecommand \@ifx [1]{%
 \ifx #1\expandafter \@firstoftwo
 \else \expandafter \@secondoftwo
 \fi
}%
\providecommand \natexlab [1]{#1}%
\providecommand \enquote  [1]{``#1''}%
\providecommand \bibnamefont  [1]{#1}%
\providecommand \bibfnamefont [1]{#1}%
\providecommand \citenamefont [1]{#1}%
\providecommand \href@noop [0]{\@secondoftwo}%
\providecommand \href [0]{\begingroup \@sanitize@url \@href}%
\providecommand \@href[1]{\@@startlink{#1}\@@href}%
\providecommand \@@href[1]{\endgroup#1\@@endlink}%
\providecommand \@sanitize@url [0]{\catcode `\\12\catcode `\$12\catcode `\&12\catcode `\#12\catcode `\^12\catcode `\_12\catcode `\%12\relax}%
\providecommand \@@startlink[1]{}%
\providecommand \@@endlink[0]{}%
\providecommand \url  [0]{\begingroup\@sanitize@url \@url }%
\providecommand \@url [1]{\endgroup\@href {#1}{\urlprefix }}%
\providecommand \urlprefix  [0]{URL }%
\providecommand \Eprint [0]{\href }%
\providecommand \doibase [0]{http://dx.doi.org/}%
\providecommand \selectlanguage [0]{\@gobble}%
\providecommand \bibinfo  [0]{\@secondoftwo}%
\providecommand \bibfield  [0]{\@secondoftwo}%
\providecommand \translation [1]{[#1]}%
\providecommand \BibitemOpen [0]{}%
\providecommand \bibitemStop [0]{}%
\providecommand \bibitemNoStop [0]{.\EOS\space}%
\providecommand \EOS [0]{\spacefactor3000\relax}%
\providecommand \BibitemShut  [1]{\csname bibitem#1\endcsname}%
\let\auto@bib@innerbib\@empty
\bibitem [{\citenamefont {Seifert}(2012)}]{seifert_2012}%
  \BibitemOpen
  \bibfield  {author} {\bibinfo {author} {\bibfnamefont {U.}~\bibnamefont {Seifert}},\ }\href {\doibase 10.1088/0034-4885/75/12/126001} {\bibfield  {journal} {\bibinfo  {journal} {Reports on Progress in Physics}\ }\textbf {\bibinfo {volume} {75}},\ \bibinfo {pages} {126001} (\bibinfo {year} {2012})}\BibitemShut {NoStop}%
\bibitem [{\citenamefont {Sekimoto}(2010)}]{sekimoto}%
  \BibitemOpen
  \bibfield  {author} {\bibinfo {author} {\bibfnamefont {K.}~\bibnamefont {Sekimoto}},\ }\href@noop {} {\emph {\bibinfo {title} {Stochastic Energetics}}}\ (\bibinfo  {publisher} {Courier Corporation},\ \bibinfo {year} {2010})\BibitemShut {NoStop}%
\bibitem [{\citenamefont {Shiraishi}(2023)}]{Shiraishi_2023_book}%
  \BibitemOpen
  \bibfield  {author} {\bibinfo {author} {\bibfnamefont {N.}~\bibnamefont {Shiraishi}},\ }\href {\doibase 10.1007/978-981-19-8186-9_3} {\emph {\bibinfo {title} {An Introduction to Stochastic Thermodynamics: From Basic to Advanced}}}\ (\bibinfo  {publisher} {Springer Nature Singapore},\ \bibinfo {address} {Singapore},\ \bibinfo {year} {2023})\ pp.\ \bibinfo {pages} {31--47}\BibitemShut {NoStop}%
\bibitem [{\citenamefont {Mohite}\ and\ \citenamefont {Rieger}(2026{\natexlab{a}})}]{ATM_2024_nr_st}%
  \BibitemOpen
  \bibfield  {author} {\bibinfo {author} {\bibfnamefont {A.~T.}\ \bibnamefont {Mohite}}\ and\ \bibinfo {author} {\bibfnamefont {H.}~\bibnamefont {Rieger}},\ }\href {\doibase 10.1103/mqpg-4l5h} {\bibfield  {journal} {\bibinfo  {journal} {Phys. Rev. E}\ }\textbf {\bibinfo {volume} {113}},\ \bibinfo {pages} {064136} (\bibinfo {year} {2026}{\natexlab{a}})}\BibitemShut {NoStop}%
\bibitem [{\citenamefont {Schnakenberg}(1976)}]{schnakenberg_1976}%
  \BibitemOpen
  \bibfield  {author} {\bibinfo {author} {\bibfnamefont {J.}~\bibnamefont {Schnakenberg}},\ }\href {\doibase 10.1103/RevModPhys.48.571} {\bibfield  {journal} {\bibinfo  {journal} {Rev. Mod. Phys.}\ }\textbf {\bibinfo {volume} {48}},\ \bibinfo {pages} {571} (\bibinfo {year} {1976})}\BibitemShut {NoStop}%
\bibitem [{\citenamefont {{Bochkov}}\ and\ \citenamefont {{Kuzovlev}}(1977)}]{Bochkov_1977}%
  \BibitemOpen
  \bibfield  {author} {\bibinfo {author} {\bibfnamefont {G.~N.}\ \bibnamefont {{Bochkov}}}\ and\ \bibinfo {author} {\bibfnamefont {I.~E.}\ \bibnamefont {{Kuzovlev}}},\ }\href@noop {} {\bibfield  {journal} {\bibinfo  {journal} {Zhurnal Eksperimentalnoi i Teoreticheskoi Fiziki}\ }\textbf {\bibinfo {volume} {72}},\ \bibinfo {pages} {238} (\bibinfo {year} {1977})}\BibitemShut {NoStop}%
\bibitem [{\citenamefont {{Bochkov}}\ and\ \citenamefont {{Kuzovlev}}(1979)}]{Bochkov_1979}%
  \BibitemOpen
  \bibfield  {author} {\bibinfo {author} {\bibfnamefont {G.~N.}\ \bibnamefont {{Bochkov}}}\ and\ \bibinfo {author} {\bibfnamefont {Y.~E.}\ \bibnamefont {{Kuzovlev}}},\ }\href@noop {} {\bibfield  {journal} {\bibinfo  {journal} {Soviet Journal of Experimental and Theoretical Physics}\ }\textbf {\bibinfo {volume} {49}},\ \bibinfo {pages} {543} (\bibinfo {year} {1979})}\BibitemShut {NoStop}%
\bibitem [{\citenamefont {Evans}\ \emph {et~al.}(1993)\citenamefont {Evans}, \citenamefont {Cohen},\ and\ \citenamefont {Morriss}}]{Evans_1993}%
  \BibitemOpen
  \bibfield  {author} {\bibinfo {author} {\bibfnamefont {D.~J.}\ \bibnamefont {Evans}}, \bibinfo {author} {\bibfnamefont {E.~G.~D.}\ \bibnamefont {Cohen}}, \ and\ \bibinfo {author} {\bibfnamefont {G.~P.}\ \bibnamefont {Morriss}},\ }\href {\doibase 10.1103/PhysRevLett.71.2401} {\bibfield  {journal} {\bibinfo  {journal} {Phys. Rev. Lett.}\ }\textbf {\bibinfo {volume} {71}},\ \bibinfo {pages} {2401} (\bibinfo {year} {1993})}\BibitemShut {NoStop}%
\bibitem [{\citenamefont {Evans}\ and\ \citenamefont {Searles}(1994)}]{Evans_1994}%
  \BibitemOpen
  \bibfield  {author} {\bibinfo {author} {\bibfnamefont {D.~J.}\ \bibnamefont {Evans}}\ and\ \bibinfo {author} {\bibfnamefont {D.~J.}\ \bibnamefont {Searles}},\ }\href {\doibase 10.1103/PhysRevE.50.1645} {\bibfield  {journal} {\bibinfo  {journal} {Phys. Rev. E}\ }\textbf {\bibinfo {volume} {50}},\ \bibinfo {pages} {1645} (\bibinfo {year} {1994})}\BibitemShut {NoStop}%
\bibitem [{\citenamefont {Jarzynski}(1997{\natexlab{a}})}]{Jarzynski_1997}%
  \BibitemOpen
  \bibfield  {author} {\bibinfo {author} {\bibfnamefont {C.}~\bibnamefont {Jarzynski}},\ }\href {\doibase 10.1103/PhysRevLett.78.2690} {\bibfield  {journal} {\bibinfo  {journal} {Phys. Rev. Lett.}\ }\textbf {\bibinfo {volume} {78}},\ \bibinfo {pages} {2690} (\bibinfo {year} {1997}{\natexlab{a}})}\BibitemShut {NoStop}%
\bibitem [{\citenamefont {Jarzynski}(1997{\natexlab{b}})}]{Jarzynski_1997_pre}%
  \BibitemOpen
  \bibfield  {author} {\bibinfo {author} {\bibfnamefont {C.}~\bibnamefont {Jarzynski}},\ }\href {\doibase 10.1103/PhysRevE.56.5018} {\bibfield  {journal} {\bibinfo  {journal} {Phys. Rev. E}\ }\textbf {\bibinfo {volume} {56}},\ \bibinfo {pages} {5018} (\bibinfo {year} {1997}{\natexlab{b}})}\BibitemShut {NoStop}%
\bibitem [{\citenamefont {Crooks}(1999)}]{Crooks_1999}%
  \BibitemOpen
  \bibfield  {author} {\bibinfo {author} {\bibfnamefont {G.~E.}\ \bibnamefont {Crooks}},\ }\href {\doibase 10.1103/PhysRevE.60.2721} {\bibfield  {journal} {\bibinfo  {journal} {Phys. Rev. E}\ }\textbf {\bibinfo {volume} {60}},\ \bibinfo {pages} {2721} (\bibinfo {year} {1999})}\BibitemShut {NoStop}%
\bibitem [{\citenamefont {Tasaki}(2000)}]{Tasaki_2000}%
  \BibitemOpen
  \bibfield  {author} {\bibinfo {author} {\bibfnamefont {H.}~\bibnamefont {Tasaki}},\ }\href@noop {} {} (\bibinfo {year} {2000}),\ \Eprint {http://arxiv.org/abs/cond-mat/0009244} {arXiv:cond-mat/0009244 [cond-mat.stat-mech]} \BibitemShut {NoStop}%
\bibitem [{\citenamefont {Crooks}(2000)}]{Crooks_2000}%
  \BibitemOpen
  \bibfield  {author} {\bibinfo {author} {\bibfnamefont {G.~E.}\ \bibnamefont {Crooks}},\ }\href {\doibase 10.1103/PhysRevE.61.2361} {\bibfield  {journal} {\bibinfo  {journal} {Phys. Rev. E}\ }\textbf {\bibinfo {volume} {61}},\ \bibinfo {pages} {2361} (\bibinfo {year} {2000})}\BibitemShut {NoStop}%
\bibitem [{\citenamefont {Maes}\ and\ \citenamefont {Neto{\v{c}}n{\'y}}(2003)}]{Maes_2003}%
  \BibitemOpen
  \bibfield  {author} {\bibinfo {author} {\bibfnamefont {C.}~\bibnamefont {Maes}}\ and\ \bibinfo {author} {\bibfnamefont {K.}~\bibnamefont {Neto{\v{c}}n{\'y}}},\ }\href {https://doi.org/10.1023/A:1021026930129} {\bibfield  {journal} {\bibinfo  {journal} {Journal of Statistical Physics}\ }\textbf {\bibinfo {volume} {110}},\ \bibinfo {pages} {269} (\bibinfo {year} {2003})}\BibitemShut {NoStop}%
\bibitem [{\citenamefont {Gallavotti}\ and\ \citenamefont {Cohen}(1995)}]{Gallavotti_1995}%
  \BibitemOpen
  \bibfield  {author} {\bibinfo {author} {\bibfnamefont {G.}~\bibnamefont {Gallavotti}}\ and\ \bibinfo {author} {\bibfnamefont {E.~G.~D.}\ \bibnamefont {Cohen}},\ }\href {\doibase 10.1103/PhysRevLett.74.2694} {\bibfield  {journal} {\bibinfo  {journal} {Phys. Rev. Lett.}\ }\textbf {\bibinfo {volume} {74}},\ \bibinfo {pages} {2694} (\bibinfo {year} {1995})}\BibitemShut {NoStop}%
\bibitem [{\citenamefont {Lebowitz}\ and\ \citenamefont {Spohn}(1999)}]{Lebowitz_1999}%
  \BibitemOpen
  \bibfield  {author} {\bibinfo {author} {\bibfnamefont {J.~L.}\ \bibnamefont {Lebowitz}}\ and\ \bibinfo {author} {\bibfnamefont {H.}~\bibnamefont {Spohn}},\ }\href {\doibase 10.1023/A:1004589714161} {\bibfield  {journal} {\bibinfo  {journal} {Journal of Statistical Physics}\ }\textbf {\bibinfo {volume} {95}},\ \bibinfo {pages} {333} (\bibinfo {year} {1999})}\BibitemShut {NoStop}%
\bibitem [{\citenamefont {Kurchan}(1998)}]{Kurchan_1998}%
  \BibitemOpen
  \bibfield  {author} {\bibinfo {author} {\bibfnamefont {J.}~\bibnamefont {Kurchan}},\ }\href {\doibase 10.1088/0305-4470/31/16/003} {\bibfield  {journal} {\bibinfo  {journal} {Journal of Physics A: Mathematical and General}\ }\textbf {\bibinfo {volume} {31}},\ \bibinfo {pages} {3719} (\bibinfo {year} {1998})}\BibitemShut {NoStop}%
\bibitem [{\citenamefont {Sekimoto}(1997)}]{Sekimoto_1997}%
  \BibitemOpen
  \bibfield  {author} {\bibinfo {author} {\bibfnamefont {K.}~\bibnamefont {Sekimoto}},\ }\href {\doibase 10.1143/JPSJ.66.1234} {\bibfield  {journal} {\bibinfo  {journal} {Journal of the Physical Society of Japan}\ }\textbf {\bibinfo {volume} {66}},\ \bibinfo {pages} {1234} (\bibinfo {year} {1997})}\BibitemShut {NoStop}%
\bibitem [{\citenamefont {Sekimoto}(1998)}]{Sekimoto_1998}%
  \BibitemOpen
  \bibfield  {author} {\bibinfo {author} {\bibfnamefont {K.}~\bibnamefont {Sekimoto}},\ }\href {\doibase 10.1143/PTPS.130.17} {\bibfield  {journal} {\bibinfo  {journal} {Progress of Theoretical Physics Supplement}\ }\textbf {\bibinfo {volume} {130}},\ \bibinfo {pages} {17} (\bibinfo {year} {1998})}\BibitemShut {NoStop}%
\bibitem [{\citenamefont {Seifert}(2005)}]{Seifert_2005}%
  \BibitemOpen
  \bibfield  {author} {\bibinfo {author} {\bibfnamefont {U.}~\bibnamefont {Seifert}},\ }\href {\doibase 10.1103/PhysRevLett.95.040602} {\bibfield  {journal} {\bibinfo  {journal} {Phys. Rev. Lett.}\ }\textbf {\bibinfo {volume} {95}},\ \bibinfo {pages} {040602} (\bibinfo {year} {2005})}\BibitemShut {NoStop}%
\bibitem [{\citenamefont {Gilmore}(1985)}]{Gilmore_1985}%
  \BibitemOpen
  \bibfield  {author} {\bibinfo {author} {\bibfnamefont {R.}~\bibnamefont {Gilmore}},\ }\href {\doibase 10.1103/PhysRevA.31.3237} {\bibfield  {journal} {\bibinfo  {journal} {Phys. Rev. A}\ }\textbf {\bibinfo {volume} {31}},\ \bibinfo {pages} {3237} (\bibinfo {year} {1985})}\BibitemShut {NoStop}%
\bibitem [{\citenamefont {Uffink}\ and\ \citenamefont {van Lith}(1999)}]{Uffink_1999}%
  \BibitemOpen
  \bibfield  {author} {\bibinfo {author} {\bibfnamefont {J.}~\bibnamefont {Uffink}}\ and\ \bibinfo {author} {\bibfnamefont {J.}~\bibnamefont {van Lith}},\ }\href {\doibase 10.1023/A:1018811305766} {\bibfield  {journal} {\bibinfo  {journal} {Foundations of Physics}\ }\textbf {\bibinfo {volume} {29}},\ \bibinfo {pages} {655} (\bibinfo {year} {1999})}\BibitemShut {NoStop}%
\bibitem [{\citenamefont {Horowitz}\ and\ \citenamefont {Gingrich}(2020)}]{Horowitz_2020}%
  \BibitemOpen
  \bibfield  {author} {\bibinfo {author} {\bibfnamefont {J.~M.}\ \bibnamefont {Horowitz}}\ and\ \bibinfo {author} {\bibfnamefont {T.~R.}\ \bibnamefont {Gingrich}},\ }\href {\doibase 10.1038/s41567-019-0702-6} {\bibfield  {journal} {\bibinfo  {journal} {Nature Physics}\ }\textbf {\bibinfo {volume} {16}},\ \bibinfo {pages} {15} (\bibinfo {year} {2020})}\BibitemShut {NoStop}%
\bibitem [{\citenamefont {Barato}\ and\ \citenamefont {Seifert}(2015)}]{Barato_2015}%
  \BibitemOpen
  \bibfield  {author} {\bibinfo {author} {\bibfnamefont {A.~C.}\ \bibnamefont {Barato}}\ and\ \bibinfo {author} {\bibfnamefont {U.}~\bibnamefont {Seifert}},\ }\href {\doibase 10.1103/PhysRevLett.114.158101} {\bibfield  {journal} {\bibinfo  {journal} {Phys. Rev. Lett.}\ }\textbf {\bibinfo {volume} {114}},\ \bibinfo {pages} {158101} (\bibinfo {year} {2015})}\BibitemShut {NoStop}%
\bibitem [{\citenamefont {Gingrich}\ \emph {et~al.}(2016)\citenamefont {Gingrich}, \citenamefont {Horowitz}, \citenamefont {Perunov},\ and\ \citenamefont {England}}]{Gingrich_2016}%
  \BibitemOpen
  \bibfield  {author} {\bibinfo {author} {\bibfnamefont {T.~R.}\ \bibnamefont {Gingrich}}, \bibinfo {author} {\bibfnamefont {J.~M.}\ \bibnamefont {Horowitz}}, \bibinfo {author} {\bibfnamefont {N.}~\bibnamefont {Perunov}}, \ and\ \bibinfo {author} {\bibfnamefont {J.~L.}\ \bibnamefont {England}},\ }\href {\doibase 10.1103/PhysRevLett.116.120601} {\bibfield  {journal} {\bibinfo  {journal} {Phys. Rev. Lett.}\ }\textbf {\bibinfo {volume} {116}},\ \bibinfo {pages} {120601} (\bibinfo {year} {2016})}\BibitemShut {NoStop}%
\bibitem [{\citenamefont {Horowitz}\ and\ \citenamefont {Gingrich}(2017)}]{Horowitz_2017}%
  \BibitemOpen
  \bibfield  {author} {\bibinfo {author} {\bibfnamefont {J.~M.}\ \bibnamefont {Horowitz}}\ and\ \bibinfo {author} {\bibfnamefont {T.~R.}\ \bibnamefont {Gingrich}},\ }\href {\doibase 10.1103/PhysRevE.96.020103} {\bibfield  {journal} {\bibinfo  {journal} {Phys. Rev. E}\ }\textbf {\bibinfo {volume} {96}},\ \bibinfo {pages} {020103} (\bibinfo {year} {2017})}\BibitemShut {NoStop}%
\bibitem [{\citenamefont {Vo}\ \emph {et~al.}(2022)\citenamefont {Vo}, \citenamefont {Vu},\ and\ \citenamefont {Hasegawa}}]{Vo_2022}%
  \BibitemOpen
  \bibfield  {author} {\bibinfo {author} {\bibfnamefont {V.~T.}\ \bibnamefont {Vo}}, \bibinfo {author} {\bibfnamefont {T.~V.}\ \bibnamefont {Vu}}, \ and\ \bibinfo {author} {\bibfnamefont {Y.}~\bibnamefont {Hasegawa}},\ }\href {\doibase 10.1088/1751-8121/ac9099} {\bibfield  {journal} {\bibinfo  {journal} {Journal of Physics A: Mathematical and Theoretical}\ }\textbf {\bibinfo {volume} {55}},\ \bibinfo {pages} {405004} (\bibinfo {year} {2022})}\BibitemShut {NoStop}%
\bibitem [{\citenamefont {Kwon}\ \emph {et~al.}(2024)\citenamefont {Kwon}, \citenamefont {Park}, \citenamefont {Lee},\ and\ \citenamefont {Baek}}]{Kwon_2023_TUR_unified}%
  \BibitemOpen
  \bibfield  {author} {\bibinfo {author} {\bibfnamefont {E.}~\bibnamefont {Kwon}}, \bibinfo {author} {\bibfnamefont {J.-M.}\ \bibnamefont {Park}}, \bibinfo {author} {\bibfnamefont {J.~S.}\ \bibnamefont {Lee}}, \ and\ \bibinfo {author} {\bibfnamefont {Y.}~\bibnamefont {Baek}},\ }\href {\doibase 10.1103/PhysRevE.110.044131} {\bibfield  {journal} {\bibinfo  {journal} {Phys. Rev. E}\ }\textbf {\bibinfo {volume} {110}},\ \bibinfo {pages} {044131} (\bibinfo {year} {2024})}\BibitemShut {NoStop}%
\bibitem [{\citenamefont {Ito}(2018)}]{Ito_2018}%
  \BibitemOpen
  \bibfield  {author} {\bibinfo {author} {\bibfnamefont {S.}~\bibnamefont {Ito}},\ }\href {\doibase 10.1103/PhysRevLett.121.030605} {\bibfield  {journal} {\bibinfo  {journal} {Phys. Rev. Lett.}\ }\textbf {\bibinfo {volume} {121}},\ \bibinfo {pages} {030605} (\bibinfo {year} {2018})}\BibitemShut {NoStop}%
\bibitem [{\citenamefont {Vo}\ \emph {et~al.}(2020)\citenamefont {Vo}, \citenamefont {Van~Vu},\ and\ \citenamefont {Hasegawa}}]{Vo_2020}%
  \BibitemOpen
  \bibfield  {author} {\bibinfo {author} {\bibfnamefont {V.~T.}\ \bibnamefont {Vo}}, \bibinfo {author} {\bibfnamefont {T.}~\bibnamefont {Van~Vu}}, \ and\ \bibinfo {author} {\bibfnamefont {Y.}~\bibnamefont {Hasegawa}},\ }\href {\doibase 10.1103/PhysRevE.102.062132} {\bibfield  {journal} {\bibinfo  {journal} {Phys. Rev. E}\ }\textbf {\bibinfo {volume} {102}},\ \bibinfo {pages} {062132} (\bibinfo {year} {2020})}\BibitemShut {NoStop}%
\bibitem [{\citenamefont {Lee}\ \emph {et~al.}(2022)\citenamefont {Lee}, \citenamefont {Lee}, \citenamefont {Kwon},\ and\ \citenamefont {Park}}]{Lee_2022}%
  \BibitemOpen
  \bibfield  {author} {\bibinfo {author} {\bibfnamefont {J.~S.}\ \bibnamefont {Lee}}, \bibinfo {author} {\bibfnamefont {S.}~\bibnamefont {Lee}}, \bibinfo {author} {\bibfnamefont {H.}~\bibnamefont {Kwon}}, \ and\ \bibinfo {author} {\bibfnamefont {H.}~\bibnamefont {Park}},\ }\href {\doibase 10.1103/PhysRevLett.129.120603} {\bibfield  {journal} {\bibinfo  {journal} {Phys. Rev. Lett.}\ }\textbf {\bibinfo {volume} {129}},\ \bibinfo {pages} {120603} (\bibinfo {year} {2022})}\BibitemShut {NoStop}%
\bibitem [{\citenamefont {Touchette}(2009)}]{Touchette_2009}%
  \BibitemOpen
  \bibfield  {author} {\bibinfo {author} {\bibfnamefont {H.}~\bibnamefont {Touchette}},\ }\href {\doibase https://doi.org/10.1016/j.physrep.2009.05.002} {\bibfield  {journal} {\bibinfo  {journal} {Physics Reports}\ }\textbf {\bibinfo {volume} {478}},\ \bibinfo {pages} {1} (\bibinfo {year} {2009})}\BibitemShut {NoStop}%
\bibitem [{\citenamefont {Klein}\ and\ \citenamefont {Meijer}(1954)}]{Klein_1954}%
  \BibitemOpen
  \bibfield  {author} {\bibinfo {author} {\bibfnamefont {M.~J.}\ \bibnamefont {Klein}}\ and\ \bibinfo {author} {\bibfnamefont {P.~H.}\ \bibnamefont {Meijer}},\ }\href {https://link.aps.org/doi/10.1103/PhysRev.96.250} {\bibfield  {journal} {\bibinfo  {journal} {Physical Review}\ }\textbf {\bibinfo {volume} {96}},\ \bibinfo {pages} {250} (\bibinfo {year} {1954})}\BibitemShut {NoStop}%
\bibitem [{\citenamefont {Callen}(1957)}]{Callen_1957}%
  \BibitemOpen
  \bibfield  {author} {\bibinfo {author} {\bibfnamefont {H.~B.}\ \bibnamefont {Callen}},\ }\href {\doibase 10.1103/PhysRev.105.360} {\bibfield  {journal} {\bibinfo  {journal} {Phys. Rev.}\ }\textbf {\bibinfo {volume} {105}},\ \bibinfo {pages} {360} (\bibinfo {year} {1957})}\BibitemShut {NoStop}%
\bibitem [{\citenamefont {Glansdorff}\ \emph {et~al.}(1974)\citenamefont {Glansdorff}, \citenamefont {Nicolis},\ and\ \citenamefont {Prigogine}}]{Glansdorff_1971}%
  \BibitemOpen
  \bibfield  {author} {\bibinfo {author} {\bibfnamefont {P.}~\bibnamefont {Glansdorff}}, \bibinfo {author} {\bibfnamefont {G.}~\bibnamefont {Nicolis}}, \ and\ \bibinfo {author} {\bibfnamefont {I.}~\bibnamefont {Prigogine}},\ }\href {https://www.pnas.org/doi/abs/10.1073/pnas.71.1.197} {\bibfield  {journal} {\bibinfo  {journal} {Proceedings of the National Academy of Sciences}\ }\textbf {\bibinfo {volume} {71}},\ \bibinfo {pages} {197} (\bibinfo {year} {1974})}\BibitemShut {NoStop}%
\bibitem [{\citenamefont {Jaynes}(1980)}]{Jaynes_1980}%
  \BibitemOpen
  \bibfield  {author} {\bibinfo {author} {\bibfnamefont {E.~T.}\ \bibnamefont {Jaynes}},\ }\href {\doibase 10.1146/annurev.pc.31.100180.003051} {\bibfield  {journal} {\bibinfo  {journal} {Annual Review of Physical Chemistry}\ }\textbf {\bibinfo {volume} {31}},\ \bibinfo {pages} {579} (\bibinfo {year} {1980})}\BibitemShut {NoStop}%
\bibitem [{\citenamefont {Struchtrup}\ and\ \citenamefont {Weiss}(1998)}]{Struchtrup_1998}%
  \BibitemOpen
  \bibfield  {author} {\bibinfo {author} {\bibfnamefont {H.}~\bibnamefont {Struchtrup}}\ and\ \bibinfo {author} {\bibfnamefont {W.}~\bibnamefont {Weiss}},\ }\href {\doibase 10.1103/PhysRevLett.80.5048} {\bibfield  {journal} {\bibinfo  {journal} {Phys. Rev. Lett.}\ }\textbf {\bibinfo {volume} {80}},\ \bibinfo {pages} {5048} (\bibinfo {year} {1998})}\BibitemShut {NoStop}%
\bibitem [{\citenamefont {Qian}(2002)}]{Qian_2002}%
  \BibitemOpen
  \bibfield  {author} {\bibinfo {author} {\bibfnamefont {H.}~\bibnamefont {Qian}},\ }\href {\doibase 10.1103/PhysRevE.65.021111} {\bibfield  {journal} {\bibinfo  {journal} {Phys. Rev. E}\ }\textbf {\bibinfo {volume} {65}},\ \bibinfo {pages} {021111} (\bibinfo {year} {2002})}\BibitemShut {NoStop}%
\bibitem [{\citenamefont {Evans}(2004{\natexlab{a}})}]{Evans_2004}%
  \BibitemOpen
  \bibfield  {author} {\bibinfo {author} {\bibfnamefont {R.~M.~L.}\ \bibnamefont {Evans}},\ }\href {\doibase 10.1103/PhysRevLett.92.150601} {\bibfield  {journal} {\bibinfo  {journal} {Phys. Rev. Lett.}\ }\textbf {\bibinfo {volume} {92}},\ \bibinfo {pages} {150601} (\bibinfo {year} {2004}{\natexlab{a}})}\BibitemShut {NoStop}%
\bibitem [{\citenamefont {Evans}(2004{\natexlab{b}})}]{Evans_2005}%
  \BibitemOpen
  \bibfield  {author} {\bibinfo {author} {\bibfnamefont {R.~M.~L.}\ \bibnamefont {Evans}},\ }\href {\doibase 10.1088/0305-4470/38/2/001} {\bibfield  {journal} {\bibinfo  {journal} {Journal of Physics A: Mathematical and General}\ }\textbf {\bibinfo {volume} {38}},\ \bibinfo {pages} {293} (\bibinfo {year} {2004}{\natexlab{b}})}\BibitemShut {NoStop}%
\bibitem [{\citenamefont {Lecomte}\ \emph {et~al.}(2005)\citenamefont {Lecomte}, \citenamefont {Appert-Rolland},\ and\ \citenamefont {van Wijland}}]{Lecomte_2005}%
  \BibitemOpen
  \bibfield  {author} {\bibinfo {author} {\bibfnamefont {V.}~\bibnamefont {Lecomte}}, \bibinfo {author} {\bibfnamefont {C.}~\bibnamefont {Appert-Rolland}}, \ and\ \bibinfo {author} {\bibfnamefont {F.}~\bibnamefont {van Wijland}},\ }\href {\doibase 10.1103/PhysRevLett.95.010601} {\bibfield  {journal} {\bibinfo  {journal} {Phys. Rev. Lett.}\ }\textbf {\bibinfo {volume} {95}},\ \bibinfo {pages} {010601} (\bibinfo {year} {2005})}\BibitemShut {NoStop}%
\bibitem [{\citenamefont {Martyushev}\ and\ \citenamefont {Seleznev}(2006)}]{Martyushev_2006}%
  \BibitemOpen
  \bibfield  {author} {\bibinfo {author} {\bibfnamefont {L.}~\bibnamefont {Martyushev}}\ and\ \bibinfo {author} {\bibfnamefont {V.}~\bibnamefont {Seleznev}},\ }\href {\doibase https://doi.org/10.1016/j.physrep.2005.12.001} {\bibfield  {journal} {\bibinfo  {journal} {Physics Reports}\ }\textbf {\bibinfo {volume} {426}},\ \bibinfo {pages} {1} (\bibinfo {year} {2006})}\BibitemShut {NoStop}%
\bibitem [{\citenamefont {Wang}(2006)}]{Wang_2006}%
  \BibitemOpen
  \bibfield  {author} {\bibinfo {author} {\bibfnamefont {Q.}~\bibnamefont {Wang}},\ }\href@noop {} {\bibfield  {journal} {\bibinfo  {journal} {Astrophysics and Space Science}\ }\textbf {\bibinfo {volume} {305}},\ \bibinfo {pages} {273} (\bibinfo {year} {2006})}\BibitemShut {NoStop}%
\bibitem [{\citenamefont {Bruers}\ \emph {et~al.}(2007)\citenamefont {Bruers}, \citenamefont {Maes},\ and\ \citenamefont {Neto{\v{c}}n{\'y}}}]{Bruers_2007}%
  \BibitemOpen
  \bibfield  {author} {\bibinfo {author} {\bibfnamefont {S.}~\bibnamefont {Bruers}}, \bibinfo {author} {\bibfnamefont {C.}~\bibnamefont {Maes}}, \ and\ \bibinfo {author} {\bibfnamefont {K.}~\bibnamefont {Neto{\v{c}}n{\'y}}},\ }\href {\doibase 10.1007/s10955-007-9412-z} {\bibfield  {journal} {\bibinfo  {journal} {Journal of Statistical Physics}\ }\textbf {\bibinfo {volume} {129}},\ \bibinfo {pages} {725} (\bibinfo {year} {2007})}\BibitemShut {NoStop}%
\bibitem [{\citenamefont {Bruers}(2007)}]{Bruers_2007_maxEP_minEP}%
  \BibitemOpen
  \bibfield  {author} {\bibinfo {author} {\bibfnamefont {S.}~\bibnamefont {Bruers}},\ }\href {\doibase 10.1088/1751-8113/40/27/003} {\bibfield  {journal} {\bibinfo  {journal} {Journal of Physics A: Mathematical and Theoretical}\ }\textbf {\bibinfo {volume} {40}},\ \bibinfo {pages} {7441} (\bibinfo {year} {2007})}\BibitemShut {NoStop}%
\bibitem [{\citenamefont {Lecomte}\ \emph {et~al.}(2007)\citenamefont {Lecomte}, \citenamefont {Appert-Rolland},\ and\ \citenamefont {van Wijland}}]{Lecomte_2007}%
  \BibitemOpen
  \bibfield  {author} {\bibinfo {author} {\bibfnamefont {V.}~\bibnamefont {Lecomte}}, \bibinfo {author} {\bibfnamefont {C.}~\bibnamefont {Appert-Rolland}}, \ and\ \bibinfo {author} {\bibfnamefont {F.}~\bibnamefont {van Wijland}},\ }\href {\doibase 10.1007/s10955-006-9254-0} {\bibfield  {journal} {\bibinfo  {journal} {Journal of Statistical Physics}\ }\textbf {\bibinfo {volume} {127}},\ \bibinfo {pages} {51} (\bibinfo {year} {2007})}\BibitemShut {NoStop}%
\bibitem [{\citenamefont {Yoshida}\ and\ \citenamefont {Mahajan}(2008)}]{Yoshida_2008}%
  \BibitemOpen
  \bibfield  {author} {\bibinfo {author} {\bibfnamefont {Z.}~\bibnamefont {Yoshida}}\ and\ \bibinfo {author} {\bibfnamefont {S.~M.}\ \bibnamefont {Mahajan}},\ }\href {https://doi.org/10.1063/1.2890189} {\bibfield  {journal} {\bibinfo  {journal} {Physics of Plasmas}\ }\textbf {\bibinfo {volume} {15}},\ \bibinfo {pages} {032307} (\bibinfo {year} {2008})}\BibitemShut {NoStop}%
\bibitem [{\citenamefont {Martyushev}(2010)}]{Martyushev_2010}%
  \BibitemOpen
  \bibfield  {author} {\bibinfo {author} {\bibfnamefont {L.~M.}\ \bibnamefont {Martyushev}},\ }\href {https://royalsocietypublishing.org/doi/abs/10.1098/rstb.2009.0295} {\bibfield  {journal} {\bibinfo  {journal} {Philosophical Transactions of the Royal Society B: Biological Sciences}\ }\textbf {\bibinfo {volume} {365}},\ \bibinfo {pages} {1333} (\bibinfo {year} {2010})}\BibitemShut {NoStop}%
\bibitem [{\citenamefont {Baule}\ and\ \citenamefont {Evans}(2010)}]{Baule_2010}%
  \BibitemOpen
  \bibfield  {author} {\bibinfo {author} {\bibfnamefont {A.}~\bibnamefont {Baule}}\ and\ \bibinfo {author} {\bibfnamefont {R.~M.~L.}\ \bibnamefont {Evans}},\ }\href {\doibase 10.1088/1742-5468/2010/03/P03030} {\bibfield  {journal} {\bibinfo  {journal} {Journal of Statistical Mechanics: Theory and Experiment}\ }\textbf {\bibinfo {volume} {2010}},\ \bibinfo {pages} {P03030} (\bibinfo {year} {2010})}\BibitemShut {NoStop}%
\bibitem [{\citenamefont {Niven}(2010)}]{Niven_2010}%
  \BibitemOpen
  \bibfield  {author} {\bibinfo {author} {\bibfnamefont {R.~K.}\ \bibnamefont {Niven}},\ }\href {https://doi.org/10.1515/jnetdy.2010.022} {\bibfield  {journal} {\bibinfo  {journal} {Journal of Non-Equilibrium Thermodynamics}\ }\textbf {\bibinfo {volume} {35}},\ \bibinfo {pages} {347} (\bibinfo {year} {2010})}\BibitemShut {NoStop}%
\bibitem [{\citenamefont {Kawazura}\ and\ \citenamefont {Yoshida}(2010)}]{Kawazura_2010}%
  \BibitemOpen
  \bibfield  {author} {\bibinfo {author} {\bibfnamefont {Y.}~\bibnamefont {Kawazura}}\ and\ \bibinfo {author} {\bibfnamefont {Z.}~\bibnamefont {Yoshida}},\ }\href {https://link.aps.org/doi/10.1103/PhysRevE.82.066403} {\bibfield  {journal} {\bibinfo  {journal} {Phys. Rev. E}\ }\textbf {\bibinfo {volume} {82}},\ \bibinfo {pages} {066403} (\bibinfo {year} {2010})}\BibitemShut {NoStop}%
\bibitem [{\citenamefont {Doi}(2011)}]{Doi_2011}%
  \BibitemOpen
  \bibfield  {author} {\bibinfo {author} {\bibfnamefont {M.}~\bibnamefont {Doi}},\ }\href {https://dx.doi.org/10.1088/0953-8984/23/28/284118} {\bibfield  {journal} {\bibinfo  {journal} {Journal of Physics: Condensed Matter}\ }\textbf {\bibinfo {volume} {23}},\ \bibinfo {pages} {284118} (\bibinfo {year} {2011})}\BibitemShut {NoStop}%
\bibitem [{\citenamefont {Monthus}(2011)}]{Monthus_2011}%
  \BibitemOpen
  \bibfield  {author} {\bibinfo {author} {\bibfnamefont {C.}~\bibnamefont {Monthus}},\ }\href {\doibase 10.1088/1742-5468/2011/03/P03008} {\bibfield  {journal} {\bibinfo  {journal} {Journal of Statistical Mechanics: Theory and Experiment}\ }\textbf {\bibinfo {volume} {2011}},\ \bibinfo {pages} {P03008} (\bibinfo {year} {2011})}\BibitemShut {NoStop}%
\bibitem [{\citenamefont {Kawazura}\ and\ \citenamefont {Yoshida}(2012)}]{Kawazura_2012}%
  \BibitemOpen
  \bibfield  {author} {\bibinfo {author} {\bibfnamefont {Y.}~\bibnamefont {Kawazura}}\ and\ \bibinfo {author} {\bibfnamefont {Z.}~\bibnamefont {Yoshida}},\ }\href {https://doi.org/10.1063/1.3675854} {\bibfield  {journal} {\bibinfo  {journal} {Physics of Plasmas}\ }\textbf {\bibinfo {volume} {19}},\ \bibinfo {pages} {012305} (\bibinfo {year} {2012})}\BibitemShut {NoStop}%
\bibitem [{\citenamefont {Chetrite}\ and\ \citenamefont {Touchette}(2013)}]{Chetrite_2013}%
  \BibitemOpen
  \bibfield  {author} {\bibinfo {author} {\bibfnamefont {R.}~\bibnamefont {Chetrite}}\ and\ \bibinfo {author} {\bibfnamefont {H.}~\bibnamefont {Touchette}},\ }\href {\doibase 10.1103/PhysRevLett.111.120601} {\bibfield  {journal} {\bibinfo  {journal} {Phys. Rev. Lett.}\ }\textbf {\bibinfo {volume} {111}},\ \bibinfo {pages} {120601} (\bibinfo {year} {2013})}\BibitemShut {NoStop}%
\bibitem [{\citenamefont {Press\'e}\ \emph {et~al.}(2013)\citenamefont {Press\'e}, \citenamefont {Ghosh}, \citenamefont {Lee},\ and\ \citenamefont {Dill}}]{Presse_2013}%
  \BibitemOpen
  \bibfield  {author} {\bibinfo {author} {\bibfnamefont {S.}~\bibnamefont {Press\'e}}, \bibinfo {author} {\bibfnamefont {K.}~\bibnamefont {Ghosh}}, \bibinfo {author} {\bibfnamefont {J.}~\bibnamefont {Lee}}, \ and\ \bibinfo {author} {\bibfnamefont {K.~A.}\ \bibnamefont {Dill}},\ }\href {\doibase 10.1103/RevModPhys.85.1115} {\bibfield  {journal} {\bibinfo  {journal} {Rev. Mod. Phys.}\ }\textbf {\bibinfo {volume} {85}},\ \bibinfo {pages} {1115} (\bibinfo {year} {2013})}\BibitemShut {NoStop}%
\bibitem [{\citenamefont {Endres}(2017)}]{Endres_2017}%
  \BibitemOpen
  \bibfield  {author} {\bibinfo {author} {\bibfnamefont {R.~G.}\ \bibnamefont {Endres}},\ }\href {\doibase 10.1038/s41598-017-14485-8} {\bibfield  {journal} {\bibinfo  {journal} {Scientific Reports}\ }\textbf {\bibinfo {volume} {7}},\ \bibinfo {pages} {14437} (\bibinfo {year} {2017})}\BibitemShut {NoStop}%
\bibitem [{\citenamefont {Bialek}(2000)}]{Bialek_2000}%
  \BibitemOpen
  \bibfield  {author} {\bibinfo {author} {\bibfnamefont {W.}~\bibnamefont {Bialek}},\ }\href@noop {} {\enquote {\bibinfo {title} {Stability and noise in biochemical switches},}\ } (\bibinfo {year} {2000}),\ \Eprint {http://arxiv.org/abs/cond-mat/0005235} {arXiv:cond-mat/0005235 [cond-mat.soft]} \BibitemShut {NoStop}%
\bibitem [{\citenamefont {Sasai}\ and\ \citenamefont {Wolynes}(2003)}]{Sasai_2003}%
  \BibitemOpen
  \bibfield  {author} {\bibinfo {author} {\bibfnamefont {M.}~\bibnamefont {Sasai}}\ and\ \bibinfo {author} {\bibfnamefont {P.~G.}\ \bibnamefont {Wolynes}},\ }\href {https://www.pnas.org/doi/abs/10.1073/pnas.2627987100} {\bibfield  {journal} {\bibinfo  {journal} {Proceedings of the National Academy of Sciences}\ }\textbf {\bibinfo {volume} {100}},\ \bibinfo {pages} {2374} (\bibinfo {year} {2003})}\BibitemShut {NoStop}%
\bibitem [{\citenamefont {Lan}\ \emph {et~al.}(2006)\citenamefont {Lan}, \citenamefont {Wolynes},\ and\ \citenamefont {Papoian}}]{Lan_2006}%
  \BibitemOpen
  \bibfield  {author} {\bibinfo {author} {\bibfnamefont {Y.}~\bibnamefont {Lan}}, \bibinfo {author} {\bibfnamefont {P.~G.}\ \bibnamefont {Wolynes}}, \ and\ \bibinfo {author} {\bibfnamefont {G.~A.}\ \bibnamefont {Papoian}},\ }\href {https://doi.org/10.1063/1.2353835} {\bibfield  {journal} {\bibinfo  {journal} {The Journal of Chemical Physics}\ }\textbf {\bibinfo {volume} {125}},\ \bibinfo {pages} {124106} (\bibinfo {year} {2006})}\BibitemShut {NoStop}%
\bibitem [{\citenamefont {Vellela}\ and\ \citenamefont {Qian}(2009)}]{Vellela_2009}%
  \BibitemOpen
  \bibfield  {author} {\bibinfo {author} {\bibfnamefont {M.}~\bibnamefont {Vellela}}\ and\ \bibinfo {author} {\bibfnamefont {H.}~\bibnamefont {Qian}},\ }\href {\doibase 10.1098/rsif.2008.0476} {\bibfield  {journal} {\bibinfo  {journal} {Journal of The Royal Society Interface}\ }\textbf {\bibinfo {volume} {6}},\ \bibinfo {pages} {925} (\bibinfo {year} {2009})}\BibitemShut {NoStop}%
\bibitem [{\citenamefont {Prados}\ \emph {et~al.}(2011)\citenamefont {Prados}, \citenamefont {Lasanta},\ and\ \citenamefont {Hurtado}}]{Prados_2011}%
  \BibitemOpen
  \bibfield  {author} {\bibinfo {author} {\bibfnamefont {A.}~\bibnamefont {Prados}}, \bibinfo {author} {\bibfnamefont {A.}~\bibnamefont {Lasanta}}, \ and\ \bibinfo {author} {\bibfnamefont {P.~I.}\ \bibnamefont {Hurtado}},\ }\href {\doibase 10.1103/PhysRevLett.107.140601} {\bibfield  {journal} {\bibinfo  {journal} {Phys. Rev. Lett.}\ }\textbf {\bibinfo {volume} {107}},\ \bibinfo {pages} {140601} (\bibinfo {year} {2011})}\BibitemShut {NoStop}%
\bibitem [{\citenamefont {Bertini}\ \emph {et~al.}(2015)\citenamefont {Bertini}, \citenamefont {De~Sole}, \citenamefont {Gabrielli}, \citenamefont {Jona-Lasinio},\ and\ \citenamefont {Landim}}]{Bertini_2015}%
  \BibitemOpen
  \bibfield  {author} {\bibinfo {author} {\bibfnamefont {L.}~\bibnamefont {Bertini}}, \bibinfo {author} {\bibfnamefont {A.}~\bibnamefont {De~Sole}}, \bibinfo {author} {\bibfnamefont {D.}~\bibnamefont {Gabrielli}}, \bibinfo {author} {\bibfnamefont {G.}~\bibnamefont {Jona-Lasinio}}, \ and\ \bibinfo {author} {\bibfnamefont {C.}~\bibnamefont {Landim}},\ }\href {\doibase 10.1103/RevModPhys.87.593} {\bibfield  {journal} {\bibinfo  {journal} {Rev. Mod. Phys.}\ }\textbf {\bibinfo {volume} {87}},\ \bibinfo {pages} {593} (\bibinfo {year} {2015})}\BibitemShut {NoStop}%
\bibitem [{\citenamefont {Bodineau}\ and\ \citenamefont {Derrida}(2004)}]{Bodineau_2004}%
  \BibitemOpen
  \bibfield  {author} {\bibinfo {author} {\bibfnamefont {T.}~\bibnamefont {Bodineau}}\ and\ \bibinfo {author} {\bibfnamefont {B.}~\bibnamefont {Derrida}},\ }\href {\doibase 10.1103/PhysRevLett.92.180601} {\bibfield  {journal} {\bibinfo  {journal} {Phys. Rev. Lett.}\ }\textbf {\bibinfo {volume} {92}},\ \bibinfo {pages} {180601} (\bibinfo {year} {2004})}\BibitemShut {NoStop}%
\bibitem [{\citenamefont {Derrida}(2007)}]{Derrida_2007}%
  \BibitemOpen
  \bibfield  {author} {\bibinfo {author} {\bibfnamefont {B.}~\bibnamefont {Derrida}},\ }\href {\doibase 10.1088/1742-5468/2007/07/P07023} {\bibfield  {journal} {\bibinfo  {journal} {Journal of Statistical Mechanics: Theory and Experiment}\ }\textbf {\bibinfo {volume} {2007}},\ \bibinfo {pages} {P07023} (\bibinfo {year} {2007})}\BibitemShut {NoStop}%
\bibitem [{\citenamefont {Qian}\ \emph {et~al.}(2020)\citenamefont {Qian}, \citenamefont {Cheng},\ and\ \citenamefont {Yang}}]{Qian_2020}%
  \BibitemOpen
  \bibfield  {author} {\bibinfo {author} {\bibfnamefont {H.}~\bibnamefont {Qian}}, \bibinfo {author} {\bibfnamefont {Y.-C.}\ \bibnamefont {Cheng}}, \ and\ \bibinfo {author} {\bibfnamefont {Y.-J.}\ \bibnamefont {Yang}},\ }\href {\doibase 10.1209/0295-5075/131/50002} {\bibfield  {journal} {\bibinfo  {journal} {Europhysics Letters}\ }\textbf {\bibinfo {volume} {131}},\ \bibinfo {pages} {50002} (\bibinfo {year} {2020})}\BibitemShut {NoStop}%
\bibitem [{\citenamefont {Onsager}\ and\ \citenamefont {Machlup}(1953)}]{Onsager_1953}%
  \BibitemOpen
  \bibfield  {author} {\bibinfo {author} {\bibfnamefont {L.}~\bibnamefont {Onsager}}\ and\ \bibinfo {author} {\bibfnamefont {S.}~\bibnamefont {Machlup}},\ }\href {\doibase 10.1103/PhysRev.91.1505} {\bibfield  {journal} {\bibinfo  {journal} {Phys. Rev.}\ }\textbf {\bibinfo {volume} {91}},\ \bibinfo {pages} {1505} (\bibinfo {year} {1953})}\BibitemShut {NoStop}%
\bibitem [{\citenamefont {Machlup}\ and\ \citenamefont {Onsager}(1953)}]{Onsager_1953_2}%
  \BibitemOpen
  \bibfield  {author} {\bibinfo {author} {\bibfnamefont {S.}~\bibnamefont {Machlup}}\ and\ \bibinfo {author} {\bibfnamefont {L.}~\bibnamefont {Onsager}},\ }\href {\doibase 10.1103/PhysRev.91.1512} {\bibfield  {journal} {\bibinfo  {journal} {Phys. Rev.}\ }\textbf {\bibinfo {volume} {91}},\ \bibinfo {pages} {1512} (\bibinfo {year} {1953})}\BibitemShut {NoStop}%
\bibitem [{\citenamefont {Baule}\ and\ \citenamefont {Sollich}(2023)}]{Baule_2023}%
  \BibitemOpen
  \bibfield  {author} {\bibinfo {author} {\bibfnamefont {A.}~\bibnamefont {Baule}}\ and\ \bibinfo {author} {\bibfnamefont {P.}~\bibnamefont {Sollich}},\ }\href {https://doi.org/10.1038/s41598-023-30577-0} {\bibfield  {journal} {\bibinfo  {journal} {Scientific Reports}\ }\textbf {\bibinfo {volume} {13}},\ \bibinfo {pages} {3853} (\bibinfo {year} {2023})}\BibitemShut {NoStop}%
\bibitem [{\citenamefont {Huang}\ \emph {et~al.}(2025)\citenamefont {Huang}, \citenamefont {Zhou},\ and\ \citenamefont {Duan}}]{Huang_2025}%
  \BibitemOpen
  \bibfield  {author} {\bibinfo {author} {\bibfnamefont {Y.}~\bibnamefont {Huang}}, \bibinfo {author} {\bibfnamefont {X.}~\bibnamefont {Zhou}}, \ and\ \bibinfo {author} {\bibfnamefont {J.}~\bibnamefont {Duan}},\ }\href {\doibase 10.1137/24M1689259} {\bibfield  {journal} {\bibinfo  {journal} {SIAM Journal on Applied Mathematics}\ }\textbf {\bibinfo {volume} {85}},\ \bibinfo {pages} {524} (\bibinfo {year} {2025})}\BibitemShut {NoStop}%
\bibitem [{\citenamefont {Dechant}(2018)}]{Dechant_2018_multidimensional}%
  \BibitemOpen
  \bibfield  {author} {\bibinfo {author} {\bibfnamefont {A.}~\bibnamefont {Dechant}},\ }\href@noop {} {\bibfield  {journal} {\bibinfo  {journal} {Journal of Physics A: Mathematical and Theoretical}\ }\textbf {\bibinfo {volume} {52}},\ \bibinfo {pages} {035001} (\bibinfo {year} {2018})}\BibitemShut {NoStop}%
\bibitem [{\citenamefont {Hasegawa}\ and\ \citenamefont {Van~Vu}(2019)}]{Hasegawa_2019_pre}%
  \BibitemOpen
  \bibfield  {author} {\bibinfo {author} {\bibfnamefont {Y.}~\bibnamefont {Hasegawa}}\ and\ \bibinfo {author} {\bibfnamefont {T.}~\bibnamefont {Van~Vu}},\ }\href {\doibase 10.1103/PhysRevE.99.062126} {\bibfield  {journal} {\bibinfo  {journal} {Phys. Rev. E}\ }\textbf {\bibinfo {volume} {99}},\ \bibinfo {pages} {062126} (\bibinfo {year} {2019})}\BibitemShut {NoStop}%
\bibitem [{\citenamefont {Van~Vu}\ and\ \citenamefont {Hasegawa}(2020)}]{Van_Vu_2020_tur}%
  \BibitemOpen
  \bibfield  {author} {\bibinfo {author} {\bibfnamefont {T.}~\bibnamefont {Van~Vu}}\ and\ \bibinfo {author} {\bibfnamefont {Y.}~\bibnamefont {Hasegawa}},\ }\href {\doibase 10.1103/PhysRevResearch.2.013060} {\bibfield  {journal} {\bibinfo  {journal} {Phys. Rev. Res.}\ }\textbf {\bibinfo {volume} {2}},\ \bibinfo {pages} {013060} (\bibinfo {year} {2020})}\BibitemShut {NoStop}%
\bibitem [{\citenamefont {Van~Vu}\ and\ \citenamefont {Hasegawa}(2021)}]{VanVu_2021}%
  \BibitemOpen
  \bibfield  {author} {\bibinfo {author} {\bibfnamefont {T.}~\bibnamefont {Van~Vu}}\ and\ \bibinfo {author} {\bibfnamefont {Y.}~\bibnamefont {Hasegawa}},\ }\href {\doibase 10.1103/PhysRevLett.126.010601} {\bibfield  {journal} {\bibinfo  {journal} {Phys. Rev. Lett.}\ }\textbf {\bibinfo {volume} {126}},\ \bibinfo {pages} {010601} (\bibinfo {year} {2021})}\BibitemShut {NoStop}%
\bibitem [{\citenamefont {Kolchinsky}\ \emph {et~al.}(2022)\citenamefont {Kolchinsky}, \citenamefont {Dechant}, \citenamefont {Yoshimura},\ and\ \citenamefont {Ito}}]{kolchinsky_2022_information_geometry_epr}%
  \BibitemOpen
  \bibfield  {author} {\bibinfo {author} {\bibfnamefont {A.}~\bibnamefont {Kolchinsky}}, \bibinfo {author} {\bibfnamefont {A.}~\bibnamefont {Dechant}}, \bibinfo {author} {\bibfnamefont {K.}~\bibnamefont {Yoshimura}}, \ and\ \bibinfo {author} {\bibfnamefont {S.}~\bibnamefont {Ito}},\ }\href@noop {} {} (\bibinfo {year} {2022}),\ \Eprint {http://arxiv.org/abs/2206.14599} {arXiv:2206.14599 [cond-mat.stat-mech]} \BibitemShut {NoStop}%
\bibitem [{\citenamefont {Ito}\ and\ \citenamefont {Dechant}(2020)}]{Ito_2020}%
  \BibitemOpen
  \bibfield  {author} {\bibinfo {author} {\bibfnamefont {S.}~\bibnamefont {Ito}}\ and\ \bibinfo {author} {\bibfnamefont {A.}~\bibnamefont {Dechant}},\ }\href {\doibase 10.1103/PhysRevX.10.021056} {\bibfield  {journal} {\bibinfo  {journal} {Phys. Rev. X}\ }\textbf {\bibinfo {volume} {10}},\ \bibinfo {pages} {021056} (\bibinfo {year} {2020})}\BibitemShut {NoStop}%
\bibitem [{\citenamefont {Ito}(2024)}]{Ito_2024_omtp_st}%
  \BibitemOpen
  \bibfield  {author} {\bibinfo {author} {\bibfnamefont {S.}~\bibnamefont {Ito}},\ }\href {\doibase 10.1007/s41884-023-00102-3} {\bibfield  {journal} {\bibinfo  {journal} {Information Geometry}\ }\textbf {\bibinfo {volume} {7}},\ \bibinfo {pages} {441} (\bibinfo {year} {2024})}\BibitemShut {NoStop}%
\bibitem [{\citenamefont {Amari}(2016)}]{Amari_2016_book}%
  \BibitemOpen
  \bibfield  {author} {\bibinfo {author} {\bibfnamefont {S.-i.}\ \bibnamefont {Amari}},\ }\href {https://link.springer.com/book/10.1007/978-4-431-55978-8#accessibility-information} {\emph {\bibinfo {title} {Information Geometry and Its Applications}}},\ Applied Mathematical Sciences\ (\bibinfo  {publisher} {Springer Tokyo},\ \bibinfo {year} {2016})\BibitemShut {NoStop}%
\bibitem [{\citenamefont {Doi}(1976)}]{Doi_1976}%
  \BibitemOpen
  \bibfield  {author} {\bibinfo {author} {\bibfnamefont {M.}~\bibnamefont {Doi}},\ }\href {\doibase 10.1088/0305-4470/9/9/008} {\bibfield  {journal} {\bibinfo  {journal} {Journal of Physics A: Mathematical and General}\ }\textbf {\bibinfo {volume} {9}},\ \bibinfo {pages} {1465} (\bibinfo {year} {1976})}\BibitemShut {NoStop}%
\bibitem [{\citenamefont {Peliti}(1985)}]{Peliti}%
  \BibitemOpen
  \bibfield  {author} {\bibinfo {author} {\bibfnamefont {L.}~\bibnamefont {Peliti}},\ }\href {\doibase 10.1051/jphys:019850046090146900} {\bibfield  {journal} {\bibinfo  {journal} {J. Phys. France}\ }\textbf {\bibinfo {volume} {46}},\ \bibinfo {pages} {1469} (\bibinfo {year} {1985})}\BibitemShut {NoStop}%
\bibitem [{\citenamefont {Weber}\ and\ \citenamefont {Frey}(2017)}]{Weber_2017}%
  \BibitemOpen
  \bibfield  {author} {\bibinfo {author} {\bibfnamefont {M.~F.}\ \bibnamefont {Weber}}\ and\ \bibinfo {author} {\bibfnamefont {E.}~\bibnamefont {Frey}},\ }\href {\doibase 10.1088/1361-6633/aa5ae2} {\bibfield  {journal} {\bibinfo  {journal} {Reports on Progress in Physics}\ }\textbf {\bibinfo {volume} {80}},\ \bibinfo {pages} {046601} (\bibinfo {year} {2017})}\BibitemShut {NoStop}%
\bibitem [{\citenamefont {Mohite}\ and\ \citenamefont {Rieger}(2025)}]{ATM_2024_nr_cg}%
  \BibitemOpen
  \bibfield  {author} {\bibinfo {author} {\bibfnamefont {A.~T.}\ \bibnamefont {Mohite}}\ and\ \bibinfo {author} {\bibfnamefont {H.}~\bibnamefont {Rieger}},\ }\href@noop {} {} (\bibinfo {year} {2025}),\ \Eprint {http://arxiv.org/abs/2508.11430} {arXiv:2508.11430 [cond-mat.stat-mech]} \BibitemShut {NoStop}%
\bibitem [{\citenamefont {Mohite}\ and\ \citenamefont {Rieger}(2026{\natexlab{b}})}]{atm_2025_var_epr_derivation}%
  \BibitemOpen
  \bibfield  {author} {\bibinfo {author} {\bibfnamefont {A.~T.}\ \bibnamefont {Mohite}}\ and\ \bibinfo {author} {\bibfnamefont {H.}~\bibnamefont {Rieger}},\ }\href {\doibase 10.1103/fmlt-px9y} {\bibfield  {journal} {\bibinfo  {journal} {Phys. Rev. E}\ }\textbf {\bibinfo {volume} {114}},\ \bibinfo {pages} {014103} (\bibinfo {year} {2026}{\natexlab{b}})}\BibitemShut {NoStop}%
\bibitem [{\citenamefont {Mohite}\ and\ \citenamefont {Rieger}(2026{\natexlab{c}})}]{atm_2025_gftoc}%
  \BibitemOpen
  \bibfield  {author} {\bibinfo {author} {\bibfnamefont {A.~T.}\ \bibnamefont {Mohite}}\ and\ \bibinfo {author} {\bibfnamefont {H.}~\bibnamefont {Rieger}},\ }\href {\doibase 10.1103/bc7b-tfl6} {\bibfield  {journal} {\bibinfo  {journal} {Phys. Rev. Res.}\ } (\bibinfo {year} {2026}{\natexlab{c}}),\ 10.1103/bc7b-tfl6}\BibitemShut {NoStop}%
\bibitem [{Note1()}]{Note1}%
  \BibitemOpen
  \bibinfo {note} {$E$ and $F_\gamma $ are measured in units of the inverse temperature $\beta $, where we set $k_b = 1$.}\BibitemShut {Stop}%
\bibitem [{\citenamefont {Maes}(2020)}]{Maes_2020}%
  \BibitemOpen
  \bibfield  {author} {\bibinfo {author} {\bibfnamefont {C.}~\bibnamefont {Maes}},\ }\href {https://www.sciencedirect.com/science/article/pii/S0370157320300120} {\bibfield  {journal} {\bibinfo  {journal} {Physics Reports}\ }\textbf {\bibinfo {volume} {850}},\ \bibinfo {pages} {1} (\bibinfo {year} {2020})}\BibitemShut {NoStop}%
\bibitem [{Note2()}]{Note2}%
  \BibitemOpen
  \bibinfo {note} {The traffic scaled with a large deviation parameter characterizes the variance. For dynamical systems, the observation time $\tau $ \cite {Maes_2017,Maes_2020,Touchette_2009} is the relevant large deviation scaling parameter and in fluctuating hydrodynamics \cite {Bertini_2015,Touchette_2009,ATM_2024_nr_cg,ATM_2024_nr_st} it is the system volume $\protect \mathcal {V}$.}\BibitemShut {Stop}%
\bibitem [{Note3()}]{Note3}%
  \BibitemOpen
  \bibinfo {note} {By construction, the transition probability measure is assumed to satisfy the normalization constraint, $1 = \DOTSI \intop \ilimits@ \protect \mathcal {P} [\{J_\gamma , T_{\gamma }, \chi _\gamma \}] \protect \vvmathbb { D } \{ J_{\gamma } \} \protect \vvmathbb { D } \{ T_{\gamma } \} \protect \vvmathbb { D } \{ \chi _\gamma \} $. where $\protect \vvmathbb { D } \{ J_{\gamma } \} \protect \vvmathbb { D } \{ T_{\gamma } \}$ and $\protect \vvmathbb { D } \{ \chi _\gamma \}$ denote the path integral over all transition currents and noise realizations, respectively.}\BibitemShut {Stop}%
\bibitem [{\citenamefont {Andrieux}\ and\ \citenamefont {Gaspard}(2007{\natexlab{a}})}]{Andrieux_2007}%
  \BibitemOpen
  \bibfield  {author} {\bibinfo {author} {\bibfnamefont {D.}~\bibnamefont {Andrieux}}\ and\ \bibinfo {author} {\bibfnamefont {P.}~\bibnamefont {Gaspard}},\ }\href {\doibase 10.1007/s10955-006-9233-5} {\bibfield  {journal} {\bibinfo  {journal} {Journal of Statistical Physics}\ }\textbf {\bibinfo {volume} {127}},\ \bibinfo {pages} {107} (\bibinfo {year} {2007}{\natexlab{a}})}\BibitemShut {NoStop}%
\bibitem [{\citenamefont {Andrieux}\ and\ \citenamefont {Gaspard}(2007{\natexlab{b}})}]{Andrieux_2007_single_current_FT}%
  \BibitemOpen
  \bibfield  {author} {\bibinfo {author} {\bibfnamefont {D.}~\bibnamefont {Andrieux}}\ and\ \bibinfo {author} {\bibfnamefont {P.}~\bibnamefont {Gaspard}},\ }\href {\doibase 10.1016/j.crhy.2007.04.016} {\bibfield  {journal} {\bibinfo  {journal} {Comptes Rendus. Physique}\ }\textbf {\bibinfo {volume} {8}},\ \bibinfo {pages} {579} (\bibinfo {year} {2007}{\natexlab{b}})}\BibitemShut {NoStop}%
\bibitem [{\citenamefont {Salamon}\ and\ \citenamefont {Berry}(1983)}]{Salamon_1983}%
  \BibitemOpen
  \bibfield  {author} {\bibinfo {author} {\bibfnamefont {P.}~\bibnamefont {Salamon}}\ and\ \bibinfo {author} {\bibfnamefont {R.~S.}\ \bibnamefont {Berry}},\ }\href {\doibase 10.1103/PhysRevLett.51.1127} {\bibfield  {journal} {\bibinfo  {journal} {Phys. Rev. Lett.}\ }\textbf {\bibinfo {volume} {51}},\ \bibinfo {pages} {1127} (\bibinfo {year} {1983})}\BibitemShut {NoStop}%
\bibitem [{\citenamefont {{Salamon}}\ \emph {et~al.}(1985)\citenamefont {{Salamon}}, \citenamefont {{Nulton}},\ and\ \citenamefont {{Berry}}}]{Salamon_1985}%
  \BibitemOpen
  \bibfield  {author} {\bibinfo {author} {\bibfnamefont {P.}~\bibnamefont {{Salamon}}}, \bibinfo {author} {\bibfnamefont {J.~D.}\ \bibnamefont {{Nulton}}}, \ and\ \bibinfo {author} {\bibfnamefont {R.~S.}\ \bibnamefont {{Berry}}},\ }\href {\doibase 10.1063/1.448337} {\bibfield  {journal} {\bibinfo  {journal} {\jcp}\ }\textbf {\bibinfo {volume} {82}},\ \bibinfo {pages} {2433} (\bibinfo {year} {1985})}\BibitemShut {NoStop}%
\bibitem [{\citenamefont {Schl{\"o}gl}(1985)}]{Schlogl_1985}%
  \BibitemOpen
  \bibfield  {author} {\bibinfo {author} {\bibfnamefont {F.}~\bibnamefont {Schl{\"o}gl}},\ }\href {https://link.springer.com/article/10.1007/BF01328857#citeas} {\bibfield  {journal} {\bibinfo  {journal} {Zeitschrift f{\"u}r Physik B Condensed Matter}\ }\textbf {\bibinfo {volume} {59}},\ \bibinfo {pages} {449} (\bibinfo {year} {1985})}\BibitemShut {NoStop}%
\bibitem [{\citenamefont {Brody}\ and\ \citenamefont {Rivier}(1995)}]{Brody_1995}%
  \BibitemOpen
  \bibfield  {author} {\bibinfo {author} {\bibfnamefont {D.}~\bibnamefont {Brody}}\ and\ \bibinfo {author} {\bibfnamefont {N.}~\bibnamefont {Rivier}},\ }\href {https://link.aps.org/doi/10.1103/PhysRevE.51.1006} {\bibfield  {journal} {\bibinfo  {journal} {Phys. Rev. E}\ }\textbf {\bibinfo {volume} {51}},\ \bibinfo {pages} {1006} (\bibinfo {year} {1995})}\BibitemShut {NoStop}%
\bibitem [{\citenamefont {Crooks}(2007)}]{Crooks_2007}%
  \BibitemOpen
  \bibfield  {author} {\bibinfo {author} {\bibfnamefont {G.~E.}\ \bibnamefont {Crooks}},\ }\href {\doibase 10.1103/PhysRevLett.99.100602} {\bibfield  {journal} {\bibinfo  {journal} {Phys. Rev. Lett.}\ }\textbf {\bibinfo {volume} {99}},\ \bibinfo {pages} {100602} (\bibinfo {year} {2007})}\BibitemShut {NoStop}%
\bibitem [{\citenamefont {Maes}\ and\ \citenamefont {Netočný}(2008)}]{Maes_2008}%
  \BibitemOpen
  \bibfield  {author} {\bibinfo {author} {\bibfnamefont {C.}~\bibnamefont {Maes}}\ and\ \bibinfo {author} {\bibfnamefont {K.}~\bibnamefont {Netočný}},\ }\href {\doibase 10.1209/0295-5075/82/30003} {\bibfield  {journal} {\bibinfo  {journal} {Europhysics Letters}\ }\textbf {\bibinfo {volume} {82}},\ \bibinfo {pages} {30003} (\bibinfo {year} {2008})}\BibitemShut {NoStop}%
\bibitem [{\citenamefont {Maes}\ \emph {et~al.}(2009)\citenamefont {Maes}, \citenamefont {Netočný},\ and\ \citenamefont {Wynants}}]{Maes_2009}%
  \BibitemOpen
  \bibfield  {author} {\bibinfo {author} {\bibfnamefont {C.}~\bibnamefont {Maes}}, \bibinfo {author} {\bibfnamefont {K.}~\bibnamefont {Netočný}}, \ and\ \bibinfo {author} {\bibfnamefont {B.}~\bibnamefont {Wynants}},\ }\href {https://dx.doi.org/10.1088/1751-8113/42/36/365002} {\bibfield  {journal} {\bibinfo  {journal} {Journal of Physics A: Mathematical and Theoretical}\ }\textbf {\bibinfo {volume} {42}},\ \bibinfo {pages} {365002} (\bibinfo {year} {2009})}\BibitemShut {NoStop}%
\bibitem [{\citenamefont {Maes}(2017)}]{Maes_2017}%
  \BibitemOpen
  \bibfield  {author} {\bibinfo {author} {\bibfnamefont {C.}~\bibnamefont {Maes}},\ }\href {https://link.aps.org/doi/10.1103/PhysRevLett.119.160601} {\bibfield  {journal} {\bibinfo  {journal} {Phys. Rev. Lett.}\ }\textbf {\bibinfo {volume} {119}},\ \bibinfo {pages} {160601} (\bibinfo {year} {2017})}\BibitemShut {NoStop}%
\bibitem [{\citenamefont {Mielke}\ \emph {et~al.}(2014)\citenamefont {Mielke}, \citenamefont {Peletier},\ and\ \citenamefont {Renger}}]{Mielke_2014_ldp}%
  \BibitemOpen
  \bibfield  {author} {\bibinfo {author} {\bibfnamefont {A.}~\bibnamefont {Mielke}}, \bibinfo {author} {\bibfnamefont {M.~A.}\ \bibnamefont {Peletier}}, \ and\ \bibinfo {author} {\bibfnamefont {D.~R.~M.}\ \bibnamefont {Renger}},\ }\href {\doibase 10.1007/s11118-014-9418-5} {\bibfield  {journal} {\bibinfo  {journal} {Potential Analysis}\ }\textbf {\bibinfo {volume} {41}},\ \bibinfo {pages} {1293} (\bibinfo {year} {2014})}\BibitemShut {NoStop}%
\bibitem [{\citenamefont {Kobayashi}\ \emph {et~al.}(2024)\citenamefont {Kobayashi}, \citenamefont {Loutchko}, \citenamefont {Kamimura}, \citenamefont {Horiguchi},\ and\ \citenamefont {Sughiyama}}]{kobayashi_2023_information_graphs_hypergraphs}%
  \BibitemOpen
  \bibfield  {author} {\bibinfo {author} {\bibfnamefont {T.~J.}\ \bibnamefont {Kobayashi}}, \bibinfo {author} {\bibfnamefont {D.}~\bibnamefont {Loutchko}}, \bibinfo {author} {\bibfnamefont {A.}~\bibnamefont {Kamimura}}, \bibinfo {author} {\bibfnamefont {S.~A.}\ \bibnamefont {Horiguchi}}, \ and\ \bibinfo {author} {\bibfnamefont {Y.}~\bibnamefont {Sughiyama}},\ }\href {https://doi.org/10.1007/s41884-023-00125-w} {\bibfield  {journal} {\bibinfo  {journal} {Information Geometry}\ }\textbf {\bibinfo {volume} {7}},\ \bibinfo {pages} {97} (\bibinfo {year} {2024})}\BibitemShut {NoStop}%
\bibitem [{\citenamefont {Terlizzi}\ \emph {et~al.}(2024)\citenamefont {Terlizzi}, \citenamefont {Gironella}, \citenamefont {Herraez-Aguilar}, \citenamefont {Betz}, \citenamefont {Monroy}, \citenamefont {Baiesi},\ and\ \citenamefont {Ritort}}]{Terlizzi_2024}%
  \BibitemOpen
  \bibfield  {author} {\bibinfo {author} {\bibfnamefont {I.~D.}\ \bibnamefont {Terlizzi}}, \bibinfo {author} {\bibfnamefont {M.}~\bibnamefont {Gironella}}, \bibinfo {author} {\bibfnamefont {D.}~\bibnamefont {Herraez-Aguilar}}, \bibinfo {author} {\bibfnamefont {T.}~\bibnamefont {Betz}}, \bibinfo {author} {\bibfnamefont {F.}~\bibnamefont {Monroy}}, \bibinfo {author} {\bibfnamefont {M.}~\bibnamefont {Baiesi}}, \ and\ \bibinfo {author} {\bibfnamefont {F.}~\bibnamefont {Ritort}},\ }\href {https://www.science.org/doi/abs/10.1126/science.adh1823} {\bibfield  {journal} {\bibinfo  {journal} {Science}\ }\textbf {\bibinfo {volume} {383}},\ \bibinfo {pages} {971} (\bibinfo {year} {2024})}\BibitemShut {NoStop}%
\bibitem [{\citenamefont {Chetrite}\ and\ \citenamefont {Gupta}(2011)}]{Chetrite_2011}%
  \BibitemOpen
  \bibfield  {author} {\bibinfo {author} {\bibfnamefont {R.}~\bibnamefont {Chetrite}}\ and\ \bibinfo {author} {\bibfnamefont {S.}~\bibnamefont {Gupta}},\ }\href {\doibase 10.1007/s10955-011-0184-0} {\bibfield  {journal} {\bibinfo  {journal} {Journal of Statistical Physics}\ }\textbf {\bibinfo {volume} {143}},\ \bibinfo {pages} {543} (\bibinfo {year} {2011})}\BibitemShut {NoStop}%
\bibitem [{\citenamefont {Neri}\ \emph {et~al.}(2017)\citenamefont {Neri}, \citenamefont {Roldan},\ and\ \citenamefont {Juelicher}}]{Neri_2017}%
  \BibitemOpen
  \bibfield  {author} {\bibinfo {author} {\bibfnamefont {I.}~\bibnamefont {Neri}}, \bibinfo {author} {\bibfnamefont {E.}~\bibnamefont {Roldan}}, \ and\ \bibinfo {author} {\bibfnamefont {F.}~\bibnamefont {Juelicher}},\ }\href {\doibase 10.1103/PhysRevX.7.011019} {\bibfield  {journal} {\bibinfo  {journal} {Phys. Rev. X}\ }\textbf {\bibinfo {volume} {7}},\ \bibinfo {pages} {011019} (\bibinfo {year} {2017})}\BibitemShut {NoStop}%
\bibitem [{\citenamefont {Guéry-Odelin}\ \emph {et~al.}(2023)\citenamefont {Guéry-Odelin}, \citenamefont {Jarzynski}, \citenamefont {Plata}, \citenamefont {Prados},\ and\ \citenamefont {Trizac}}]{Guery_Odelin_2023}%
  \BibitemOpen
  \bibfield  {author} {\bibinfo {author} {\bibfnamefont {D.}~\bibnamefont {Guéry-Odelin}}, \bibinfo {author} {\bibfnamefont {C.}~\bibnamefont {Jarzynski}}, \bibinfo {author} {\bibfnamefont {C.~A.}\ \bibnamefont {Plata}}, \bibinfo {author} {\bibfnamefont {A.}~\bibnamefont {Prados}}, \ and\ \bibinfo {author} {\bibfnamefont {E.}~\bibnamefont {Trizac}},\ }\href {\doibase 10.1088/1361-6633/acacad} {\bibfield  {journal} {\bibinfo  {journal} {Reports on Progress in Physics}\ }\textbf {\bibinfo {volume} {86}},\ \bibinfo {pages} {035902} (\bibinfo {year} {2023})}\BibitemShut {NoStop}%
\bibitem [{\citenamefont {Jordan}\ \emph {et~al.}(1998)\citenamefont {Jordan}, \citenamefont {Kinderlehrer},\ and\ \citenamefont {Otto}}]{Jordan_1998}%
  \BibitemOpen
  \bibfield  {author} {\bibinfo {author} {\bibfnamefont {R.}~\bibnamefont {Jordan}}, \bibinfo {author} {\bibfnamefont {D.}~\bibnamefont {Kinderlehrer}}, \ and\ \bibinfo {author} {\bibfnamefont {F.}~\bibnamefont {Otto}},\ }\href {\doibase 10.1137/S0036141096303359} {\bibfield  {journal} {\bibinfo  {journal} {SIAM Journal on Mathematical Analysis}\ }\textbf {\bibinfo {volume} {29}},\ \bibinfo {pages} {1} (\bibinfo {year} {1998})}\BibitemShut {NoStop}%
\bibitem [{\citenamefont {Benamou}\ and\ \citenamefont {Brenier}(2000)}]{Benamou_2000}%
  \BibitemOpen
  \bibfield  {author} {\bibinfo {author} {\bibfnamefont {J.-D.}\ \bibnamefont {Benamou}}\ and\ \bibinfo {author} {\bibfnamefont {Y.}~\bibnamefont {Brenier}},\ }\href@noop {} {\bibfield  {journal} {\bibinfo  {journal} {Numerische Mathematik}\ }\textbf {\bibinfo {volume} {84}},\ \bibinfo {pages} {375} (\bibinfo {year} {2000})}\BibitemShut {NoStop}%
\bibitem [{\citenamefont {Aurell}\ \emph {et~al.}(2011)\citenamefont {Aurell}, \citenamefont {Mejía-Monasterio},\ and\ \citenamefont {Muratore-Ginanneschi}}]{Aurell_2011}%
  \BibitemOpen
  \bibfield  {author} {\bibinfo {author} {\bibfnamefont {E.}~\bibnamefont {Aurell}}, \bibinfo {author} {\bibfnamefont {C.}~\bibnamefont {Mejía-Monasterio}}, \ and\ \bibinfo {author} {\bibfnamefont {P.}~\bibnamefont {Muratore-Ginanneschi}},\ }\href {https://doi.org/10.1103/PhysRevLett.106.250601} {\bibfield  {journal} {\bibinfo  {journal} {Physical review letters}\ }\textbf {\bibinfo {volume} {106}},\ \bibinfo {pages} {250601} (\bibinfo {year} {2011})}\BibitemShut {NoStop}%
\bibitem [{\citenamefont {Van~Vu}\ and\ \citenamefont {Saito}(2023)}]{Van_vu_2023}%
  \BibitemOpen
  \bibfield  {author} {\bibinfo {author} {\bibfnamefont {T.}~\bibnamefont {Van~Vu}}\ and\ \bibinfo {author} {\bibfnamefont {K.}~\bibnamefont {Saito}},\ }\href {\doibase 10.1103/PhysRevX.13.011013} {\bibfield  {journal} {\bibinfo  {journal} {Phys. Rev. X}\ }\textbf {\bibinfo {volume} {13}},\ \bibinfo {pages} {011013} (\bibinfo {year} {2023})}\BibitemShut {NoStop}%
\bibitem [{\citenamefont {Sekimoto}\ and\ \citenamefont {Sasa}(1997)}]{Sekimoto_1997_slow_driving_optimal_control}%
  \BibitemOpen
  \bibfield  {author} {\bibinfo {author} {\bibfnamefont {K.}~\bibnamefont {Sekimoto}}\ and\ \bibinfo {author} {\bibfnamefont {S.-i.}\ \bibnamefont {Sasa}},\ }\href {https://www.jstage.jst.go.jp/article/jpsj/66/11/66_11_3326/_article/-char/en} {\bibfield  {journal} {\bibinfo  {journal} {Journal of the Physical Society of Japan}\ }\textbf {\bibinfo {volume} {66}},\ \bibinfo {pages} {3326} (\bibinfo {year} {1997})}\BibitemShut {NoStop}%
\bibitem [{\citenamefont {Sivak}\ and\ \citenamefont {Crooks}(2012)}]{Sivak_2012}%
  \BibitemOpen
  \bibfield  {author} {\bibinfo {author} {\bibfnamefont {D.~A.}\ \bibnamefont {Sivak}}\ and\ \bibinfo {author} {\bibfnamefont {G.~E.}\ \bibnamefont {Crooks}},\ }\href {\doibase 10.1103/PhysRevLett.108.190602} {\bibfield  {journal} {\bibinfo  {journal} {Phys. Rev. Lett.}\ }\textbf {\bibinfo {volume} {108}},\ \bibinfo {pages} {190602} (\bibinfo {year} {2012})}\BibitemShut {NoStop}%
\bibitem [{\citenamefont {Mandal}\ and\ \citenamefont {Jarzynski}(2016)}]{Mandal_2016}%
  \BibitemOpen
  \bibfield  {author} {\bibinfo {author} {\bibfnamefont {D.}~\bibnamefont {Mandal}}\ and\ \bibinfo {author} {\bibfnamefont {C.}~\bibnamefont {Jarzynski}},\ }\href {\doibase 10.1088/1742-5468/2016/06/063204} {\bibfield  {journal} {\bibinfo  {journal} {Journal of Statistical Mechanics: Theory and Experiment}\ }\textbf {\bibinfo {volume} {2016}},\ \bibinfo {pages} {063204} (\bibinfo {year} {2016})}\BibitemShut {NoStop}%
\bibitem [{\citenamefont {Li}\ \emph {et~al.}(2022)\citenamefont {Li}, \citenamefont {Chen}, \citenamefont {Sun},\ and\ \citenamefont {Dong}}]{Li_2022}%
  \BibitemOpen
  \bibfield  {author} {\bibinfo {author} {\bibfnamefont {G.}~\bibnamefont {Li}}, \bibinfo {author} {\bibfnamefont {J.-F.}\ \bibnamefont {Chen}}, \bibinfo {author} {\bibfnamefont {C.~P.}\ \bibnamefont {Sun}}, \ and\ \bibinfo {author} {\bibfnamefont {H.}~\bibnamefont {Dong}},\ }\href {\doibase 10.1103/PhysRevLett.128.230603} {\bibfield  {journal} {\bibinfo  {journal} {Phys. Rev. Lett.}\ }\textbf {\bibinfo {volume} {128}},\ \bibinfo {pages} {230603} (\bibinfo {year} {2022})}\BibitemShut {NoStop}%
\bibitem [{\citenamefont {Schmiedl}\ and\ \citenamefont {Seifert}(2007)}]{Schmiedl_2007}%
  \BibitemOpen
  \bibfield  {author} {\bibinfo {author} {\bibfnamefont {T.}~\bibnamefont {Schmiedl}}\ and\ \bibinfo {author} {\bibfnamefont {U.}~\bibnamefont {Seifert}},\ }\href {\doibase 10.1103/PhysRevLett.98.108301} {\bibfield  {journal} {\bibinfo  {journal} {Phys. Rev. Lett.}\ }\textbf {\bibinfo {volume} {98}},\ \bibinfo {pages} {108301} (\bibinfo {year} {2007})}\BibitemShut {NoStop}%
\bibitem [{\citenamefont {Chen}\ \emph {et~al.}(2019)\citenamefont {Chen}, \citenamefont {Georgiou},\ and\ \citenamefont {Tannenbaum}}]{Chen_2019_stochastic_control}%
  \BibitemOpen
  \bibfield  {author} {\bibinfo {author} {\bibfnamefont {Y.}~\bibnamefont {Chen}}, \bibinfo {author} {\bibfnamefont {T.~T.}\ \bibnamefont {Georgiou}}, \ and\ \bibinfo {author} {\bibfnamefont {A.}~\bibnamefont {Tannenbaum}},\ }\href@noop {} {\bibfield  {journal} {\bibinfo  {journal} {IEEE transactions on automatic control}\ }\textbf {\bibinfo {volume} {65}},\ \bibinfo {pages} {2979} (\bibinfo {year} {2019})}\BibitemShut {NoStop}%
\bibitem [{\citenamefont {Zhong}\ and\ \citenamefont {DeWeese}(2024)}]{Zhong_2024}%
  \BibitemOpen
  \bibfield  {author} {\bibinfo {author} {\bibfnamefont {A.}~\bibnamefont {Zhong}}\ and\ \bibinfo {author} {\bibfnamefont {M.~R.}\ \bibnamefont {DeWeese}},\ }\href {\doibase 10.1103/PhysRevLett.133.057102} {\bibfield  {journal} {\bibinfo  {journal} {Phys. Rev. Lett.}\ }\textbf {\bibinfo {volume} {133}},\ \bibinfo {pages} {057102} (\bibinfo {year} {2024})}\BibitemShut {NoStop}%
\bibitem [{\citenamefont {Ma}\ \emph {et~al.}(2020)\citenamefont {Ma}, \citenamefont {Zhai}, \citenamefont {Chen}, \citenamefont {Sun},\ and\ \citenamefont {Dong}}]{Ma_2020}%
  \BibitemOpen
  \bibfield  {author} {\bibinfo {author} {\bibfnamefont {Y.-H.}\ \bibnamefont {Ma}}, \bibinfo {author} {\bibfnamefont {R.-X.}\ \bibnamefont {Zhai}}, \bibinfo {author} {\bibfnamefont {J.}~\bibnamefont {Chen}}, \bibinfo {author} {\bibfnamefont {C.~P.}\ \bibnamefont {Sun}}, \ and\ \bibinfo {author} {\bibfnamefont {H.}~\bibnamefont {Dong}},\ }\href {\doibase 10.1103/PhysRevLett.125.210601} {\bibfield  {journal} {\bibinfo  {journal} {Phys. Rev. Lett.}\ }\textbf {\bibinfo {volume} {125}},\ \bibinfo {pages} {210601} (\bibinfo {year} {2020})}\BibitemShut {NoStop}%
\end{thebibliography}%
\vskip2cm
\end{document}